\documentclass[12pt,draftclsnofoot,onecolumn]{IEEEtran}

\usepackage{amssymb}
\usepackage{amscd}
\usepackage{amsmath}
\usepackage{epsfig}
\usepackage{amsthm}
\usepackage{graphics}
\usepackage{psfrag}
\usepackage{rotating}
\usepackage{amsmath} 
\usepackage{amsfonts}
\usepackage{url}
\usepackage{color}
\usepackage{epstopdf}
\usepackage{amsthm}
\usepackage{tikz}
\usepackage{mathtools}
\usepackage{multirow}
\usepackage[final]{pdfpages}
\usepackage{enumitem}
\usepackage{multicol}
\usepackage[nolist]{acronym}
\usepackage[font= small]{caption}
\usepackage{subcaption}
\usepackage[ruled]{algorithm2e}
\usepackage{ragged2e}
\usepackage{array}
\usepackage{multirow}
\usepackage{pgfplots}
\usepackage{pgfplotstable}
\usepackage{calrsfs}
\pgfplotsset{compat=1.12}
\usepgfplotslibrary{polar}
\newtheorem{remark}{Remark}

\DeclareMathAlphabet{\pazocal}{OMS}{zplm}{m}{n}

\newfont{\bbb}{msbm10 scaled 700}

\newfont{\bb}{msbm10 scaled 1100}
\newcommand{\CC}{\mbox{\bb C}}
\newcommand{\PP}{\mbox{\bb P}}
\newcommand{\RR}{\mbox{\bb R}}

\newcommand{\EE}{\mbox{\bb E}}

\newcommand{\av}{{\bf a}}
\newcommand{\bv}{{\bf b}}
\newcommand{\cv}{{\bf c}}

\newcommand{\fv}{{\bf f}}

\newcommand{\hv}{{\bf h}}

\newcommand{\rv}{{\bf r}}
\newcommand{\sv}{{\bf s}}

\newcommand{\uv}{{\bf u}}
\newcommand{\wv}{{\bf w}}

\newcommand{\xv}{{\bf x}}
\newcommand{\yv}{{\bf y}}

\newcommand{\onev}{{\bf 1}}

\newcommand{\Am}{{\bf A}}
\newcommand{\Bm}{{\bf B}}

\newcommand{\Fm}{{\bf F}}
\newcommand{\Gm}{{\bf G}}
\newcommand{\Hm}{{\bf H}}
\newcommand{\Id}{{\bf I}}

\newcommand{\Um}{{\bf U}}

\newcommand{\Xm}{{\bf X}}
\newcommand{\Ym}{{\bf Y}}

\newcommand{\Cc}{{\cal C}}

\newcommand{\Hc}{{\cal H}}

\newcommand{\Lc}{{\cal L}}

\newcommand{\Tc}{{\cal T}}

\newcommand{\Xc}{{\cal X}}

\newcommand{\alphav}{\hbox{\boldmath$\alpha$}}
\newcommand{\betav}{\hbox{\boldmath$\beta$}}

\newcommand{\thetav}{\hbox{\boldmath$\theta$}}

\newcommand{\Psim}{\hbox{\boldmath$\Psi$}}

\renewcommand{\arg}{{\hbox{arg}}}

\renewcommand{\Re}{{\rm Re}}
\renewcommand{\Im}{{\rm Im}}
\newcommand{\eqdef}{\stackrel{\Delta}{=}}

\newcommand{\herm}{{\sf H}}

\newcommand{\transp}{{\sf T}}

\newcommand{\RED}{\color[rgb]{1.00,0.10,0.10}}

\newcommand{\Pav}{P_{\rm avg}}
\newcommand{\Na}{N_{\rm a}}
\newcommand{\Nrf}{N_{\rm rf}}
\newcommand{\Ns}{N_{\rm s}}
\renewcommand{\H}{{\scriptscriptstyle\mathsf{H}}}
\newcommand{\T}{{\scriptscriptstyle\mathsf{T}}}

\makeatletter
\def\@tempa#1{\@xp\@tempb\meaning#1\@nil#1}
\def\@tempb#1>#2#3 #4\@nil#5{%
  \@xp\ifx\csname#3\endcsname\mathaccent
    \@tempc#4?"7777\@nil#5%
  \else
    \PackageWarningNoLine{amsmath}{%
      Unable to redefine math accent \string#5}%
  \fi
}
\def\@tempc#1"#2#3#4#5#6\@nil#7{%
  \chardef\@tempd="#3\relax\set@mathaccent\@tempd{#7}{#2}{#4#5}}

\@tempa\widehat
\makeatother

\title{Beam-Space MIMO Radar for Joint \\[-16pt] Communication and Sensing with OTFS \\[-16pt] Modulation}

\author{Saeid K. Dehkordi$^{1}$, Lorenzo Gaudio$^{2}$,  Mari Kobayashi$^{3}$, Giuseppe Caire$^{1}$, Giulio Colavolpe$^{2}$
\thanks{$^1$ Technical University of Berlin, Germany. $^2$ Department of Engineering and Architecture,
University of Parma, 43124 Parma, Italy, and the CNIT Research
Unit, 43124 Parma, Italy. $^3$ Technical University of Munich, Munich, Germany, }
\thanks{Emails: s.khalilidehkordi@tu-berlin.de,lorenzo.gaudio@studenti.unipr.it, giulio.colavolpe@unipr.it, mari.kobayashi@tum.de, caire@tu-berlin.de}
}

\begin{document}

\begin{acronym}
	\acro{AWGN}{additive white Gaussian noise}
	\acro{MIMO}{Multiple-Input Multiple-Output}
	\acro{OTFS}{Orthogonal Time Frequency Space}
	\acro{ISAC}{Integrated Sensing and Communication}
	\acro{SNR}{Signal-to-Noise Ratio}
	\acro{mmWave}{millimeter wave}
	\acro{ML}{Maximum Likelihood}
	\acro{V2X}{vehicle-to-everything}
	\acro{OFDM}{Orthogonal Frequency Division Multiplexing}
	\acro{FMCW}{Frequency Modulated Continuous Wave}
	\acro{LoS}{Line-of-Sight}
	\acro{ISFFT}{Inverse Symplectic Finite Fourier Transform}
	\acro{SFFT}{Symplectic Finite Fourier Transform}
	\acro{HPBW}{half-power beamwidth}
	\acro{ULA}{Uniform Linear Array}
	\acro{CRLB}{Cram\'er-Rao Lower Bound}
	\acro{RF}{Radio Frequency}
	\acro{BF}{beamforming}
	\acro{RMSE}{root MSE}
	\acro{AoA}{Angle of Arrival}
	\acro{ISI}{Inter-Symbol Interference}
	\acro{SI}{self-interference}
	\acro{TDD}{Time Division dDuplex}
	\acro{Tx}{transmitter}
	\acro{Rx}{receiver}
	\acro{SIC}{Successive Interference Cancellation}
	\acro{PD}{probability of detection}
	\acro{HDA}{Hybrid Digital-Analog}
	\acro{PSD}{Power Spectral Density}
	\acro{FWHM}{full width at half maximum}
	\acro{SLL}{side lobe level}
	\acro{BS}{Base Station}
    \acro{FoV}{Field of View}
    \acro{CFAR}{Constant False Alarm Rate}
    \acro{OS-CFAR}{Ordered Statistic Constant False Alarm Rate}
    \acro{PSLR}{Peak-to-Sidelobe Ratio}
    \acro{DFT}{Discrete Fourier Transform}
\end{acronym}

\maketitle

\vspace{-1.5cm} 

\begin{abstract}
Motivated by automotive applications, we consider joint radar sensing and data communication for a system operating at \ac{mmWave} frequency bands, 
where a \ac{BS} is equipped with a co-located radar receiver and 
sends data using the \ac{OTFS} modulation format. 
We consider two distinct modes of operation.  
In \textit{Discovery} mode, a single common data stream is broadcast over a wide angular sector. The radar receiver must detect the presence of not yet acquired targets
and perform coarse estimation of their parameters (angle of arrival, range, and velocity).
In \textit{Tracking} mode, the \ac{BS} transmits multiple individual data streams to already acquired users via beamforming, while the radar receiver performs accurate estimation of the aforementioned parameters. 
Due to hardware complexity and power consumption constraints, we consider a hybrid digital-analog architecture where the number of RF chains and A/D converters is significantly smaller than the number of antenna array elements. 
In this case, a direct application of the conventional MIMO radar approach is not possible. Consequently, we advocate a {\em beam-space approach} where the vector observation at the radar receiver is obtained through a RF-domain beamforming matrix operating the dimensionality reduction from antennas to RF chains.  
Under this setup, we propose a likelihood function-based scheme to perform joint target detection and parameter estimation in Discovery, and high-resolution parameter estimation in Tracking mode, respectively. Our numerical results demonstrate that the proposed approach is able to reliably detect multiple targets while closely approaching the \ac{CRLB} of the corresponding parameter estimation problem.
\end{abstract}

%\vspace{-1cm}


%%%%%%%%%%%%%%%%%%%%%%%%%%%%%%%%%%%%%%%%%%%%%%%%%%%%%%%%%%%%%%%%%%%%%%%%%%%%%%%%%%%%%%%%%%%%%%%%%%%%%%
\section{Introduction}\label{sec:Introduction}

In \ac{mmWave} communications, it is crucial to compensate the high isotropic path-loss  with a highly directional \ac{BF} gain \cite{mmWSurvey}. This requires fast and accurate initial beam acquisition, which must be established before reliable data transmission can take place 
(see, e.g., \cite{song2019fully,song2018scalable} and references therein). 
For mobile applications, e.g., a \ac{BS} operating as a road-side infrastructure node and communicating with moving vehicles,  beam acquisition for new not yet connected vehicles entering the cell is particularly challenging.  Furthermore, beam tracking is necessary for the already connected users in order to follow their motion \cite{Yuan_BP,LiuCaire, DehkordiTracking}.
In this paper, we consider an enhanced \ac{BS}, equipped with a co-located radar receiver. While transmitting a data modulated signal in the downlink, the \ac{BS} uses the backscatter signals in order to detect unknown users and estimate parameters of already acquired users. We assume that the user terminals (i.e., targets) are large metal objects with a good radar cross-section such as vehicles  \cite{dokhanchi2019mmwave,zheng2019radar,hassanien2019dual}. As an extension of our previous works \cite{gaudio2020effectiveness}, this paper studies the joint target detection and parameter estimation problem at the enhanced \ac{BS} using \ac{OTFS}, a multi-carrier modulation proposed in \cite{hadani2017orthogonal} and already studied in different \ac{MIMO} configurations (see, e.g., \cite{shen2019channel,ramachandran2018mimoOTFS}). 

Motivated by the need for low-latency initial beam acquisition and accurate beam tracking mentioned above, we consider two distinct modes. 
In the first one, referred to as \textit{Discovery} mode, a single \ac{OTFS} modulated signal (e.g., some cell-dependent beacon/control signal) is broadcast over a wide angular sector and the goal of the radar receiver is to detect the presence of targets (vehicles) that are not yet acquired, as well as estimating their relevant parameters (angle of arrival, range, and velocity).  In  the second one, referred to as \textit{Tracking} mode, the \ac{BS} sends multiple individually beamformed OTFS modulated data streams to already acquired users, and the goal of the radar receiver is to provide high-resolution estimates of the users' parameters. For both cases, the output of the radar detector/estimator can be fed into suitable algorithms for beam acquisition \cite{BeamRefinement} and beam tracking (e.g., see \cite{Yuan_BP}). In this work, we are not concerned with the specific algorithms for initial acquisition and tracking and focus only on the target detection and parameter estimation problems in the two modes of interest.  

%%%%%%%%%%%%%%%%%%%%%%%%%%%%%%%%%%%%%%%%%%
There are basically two ways to estimate the angular position of a target: either using a highly directional Tx/Rx beam \cite{patole2017automotive} and scanning different angular ``bins'' in successive slots, or exploring 
{\em simultaneously} the whole angle domain with a wide angle Tx beam and using an antenna array observation at the Rx side, 
exploiting some array processing technique for \ac{AoA} estimation. 
The latter approach is usually referred to as \ac{MIMO} radar (see, e.g., \cite{li2008mimo}). 
The advantage of \ac{MIMO} radar is that the \ac{AoA} estimation can be accomplished ``off the grid'' by some super-resolution estimation technique \cite{richards2014fundamentals}, and therefore does not suffer from the discretization error in the angle domain. On the other hand, wide-angle illumination and 
array reception do not provide directional \ac{BF} gain, which may be a problem when the \ac{SNR} at the radar receiver is low. Furthermore, demodulating and sampling a large number of antennas tightly integrated in a monolithic structure is highly impractical 
in terms of chip area and power consumption/heat management, especially when the data signal has a large bandwidth as in \ac{mmWave} applications. 
While this problem does not exist in standard automotive radar based on specifically designed waveforms such as \ac{FMCW},  that are locally narrowband \cite{patole2017automotive}, it is definitely one of the major technological hurdles for digitally modulated waveforms at \ac{mmWave}. Therefore, its solution represents 
a critical step in order to enable joint communication and sensing, which is recognized as an important  trend 
in 6th generation wireless systems \cite{ISAC_Survey, Masouros_Survey}.

%%%%%%%%%%%%%%%%%%%%%%%%%%%%%%%%%%%%%%%%%%%%%%%%%%%%%%%%%%%%%%%%%%%%%%%%%%%%%%%%%%%%%%%%%%%%%%%%%

To address this problem, this paper considers 
a novel \ac{MIMO} radar approach in the {\em beam-space domain}, 
where the radar receiver  employs a \ac{HDA} architecture, widely proposed for \ac{mmWave} frequency bands (see, e.g., \cite{song2019fully,chen2018doa} and references therein). In this case, the number of 
\ac{RF} chains ($\Nrf$), implementing demodulation from RF to baseband and A/D conversion, 
is much smaller than the number of antenna array elements ($\Na$), and the dimensionality reduction from the $\Na$ antenna elements to the $\Nrf$ antenna ports is operated by a {\em reduction matrix} 
implemented in the analog RF domain. 
The columns of the reduction matrix define the coefficients of $\Nrf$ RF beamforming vectors. 
The role of such beamformers is twofold. On the one hand, they aim to ``cover'' sufficiently well 
the desired \ac{AoA} domain. On the other hand, they should provide sufficient 
\ac{BF} gain in order to achieve a good operating SNR of the radar receiver.
%%%%%%%%%%%%%%%%%%%%%%%%%%%%%%%%%%%%%%%%%%%%%%
The main contributions of this paper are summarized as follows:
\begin{enumerate}
\item We propose the use of multi-block processing with a randomized ensemble of carefully designed  
reduction matrices across the blocks to exploit the sparsity of \ac{mmWave} channels in the beam-space domain.  
%in analogy to compressed sensing \cite{haghighatshoar2017massive,haghighatshoar2018low}, 
We demonstrate the superiority of the proposed \ac{BF} codebook with respect to baseline choices, in particular, when the reduction matrices are constructed from \ac{DFT} beamforming vectors or from antenna selection. 

\item For Discovery mode, we propose a sequential target detection, parameter estimation, and \ac{SIC} that at each detection step performs a likelihood ratio threshold test. 
\ac{SIC} is used to avoid the masking effect of strong targets on weaker targets 
located at different ranges from the \ac{BS} and not sufficiently separated in the Doppler-delay-\ac{AoA} domain (near-far effect). The proposed scheme achieves multiple target detection over
a relatively wide \ac{FoV} and ranges comparable to Short/Medium range
automotive radar systems \cite{Auto_radars, 6G_Wymeersch}. Furthermore, the range can be extended by increasing the number of processed blocks. 

\item For Tracking mode, we propose a \ac{ML}-based scheme providing 
high-resolution (off-grid) estimation of \ac{AoA}, delay, and Doppler parameters for multiple targets. 
We demonstrate through simulations that the parameter estimation performance approaches very closely the \ac{CRLB} when the receiver \ac{SNR} is not too small. In this mode, since the already acquired targets are individually illuminated by highly directive (narrow) beams, we obtain good parameter estimation accuracy for ranges comparable to Medium/Long range automotive radars \cite{Auto_radars}. 
\end{enumerate}
The rest of the paper is organized as follows. In Section \ref{sec:phy-model} we define the system model and review OTFS modulation introducing the necessary notations. In Section \ref{sec:BeamDesign} we present the design of the proposed beamforming codebook for multi-block detection/estimation in the beam-space domain. Section  \ref{sec:Joint-Detection-Param-Estimation} discusses the details of the 
target detection and parameter estimation schemes. Section \ref{sec:Numerical-Results} presents numerical simulation results and comparisons to alternative and more conventional MIMO radar schemes. 
Finally, Section \ref{sec:Conclusions} provides our concluding remarks. 
The details of the \ac{BF} design algorithm are presented in Appendix \ref{appendix:BF}  
%the derivation of the \ac{CRLB} is given in Appendix \ref{app:CRLB}, 
and the method to calculate the adaptive threshold for the target detection scheme is given in Appendix \ref{CFAR}. We adopt the following notations.$(\cdot)^\transp$ denotes the transpose operation.  $(\cdot)^\herm$ denotes the Hermitian (conjugate and transpose) operation. $\left|x\right|$ denotes the absolute value of $x$ if $x\in\RR$ while $|\Xc|$ denotes the cardinality of a set $\Xc$.  $\|\xv\|$ denotes the $\ell_2$-norm of a complex or real vector $\xv$. 
$\Id_m$ denotes the $m \times m$  identity matrix. We let $[n]=\{1, \dots, n\}$ and $[0:n]=\{0, 1, \dots, n\}$ for a positive integer $n$.

%%%%%%%%%%%%%%%%%%%%%%%%%%%%%%%%%%%%%%%%%%%%%%%%%%%%%%%%%%%%%%%%%%%%%%%%%%%%%
\section{System model}\label{sec:phy-model}

\subsection{Backscatter Channel Model}\label{subsection:PhysicalModel}

We consider a system operating over a channel with carrier frequency $f_c$ and bandwidth $W$ sufficiently smaller than $f_c$, such that the narrowband array response assumptions holds \cite{VanTrees,Rotman}.\footnote{In particular, this means that the baseband array response vector is essentially invariant with frequency in the interval $[-W/2, W/2]$.}
We consider a \ac{BS} transmitter equipped with $\Nrf$ Tx RF chains driving an antenna array with $\Na$ elements, and a radar receiver co-located with the \ac{BS}. 
For simplicity of exposition, we assume that the Tx array and the Rx radar array coincide and that the Tx and Rx signals are separated by some full-duplex processing.\footnote{Full-duplex operations can be achieved with sufficient isolation between the transmitter and the (radar) detector and possibly interference analog pre-cancellation in order to prevent the (radar) detector saturation \cite{sabharwal2014band,wentworth2007applied,duarte2010fullduplexWireless}.}
However, all our results can be easily generalized to the case where the two arrays are different and sufficiently spatially separated such that the Tx signal does not saturate the radar Rx front-end. 
%No phase coherence between the Tx and the Rx array is required, so that the two arrays may be separated by several wavelengths. 

We consider a point target model, such that each target is characterized by its \ac{LoS} path only. This model is widely used in the literature (e.g., see \cite{kumari2018ieee,nguyen2017delay,grossi2018opportunisticRadar})
and in our case it can be justified by our motivating scenario of a 
BS operating as road-side infrastructure node and communicating with moving vehicles on the road. 
%In addition, \ac{mmWave} channels are characterized by large isotropic attenuation, 
%such that multipath components are typically much weaker than the \ac{LoS} and disappear below the noise floor after reflection, in particular for the backscatter channel seen by the radar receiver.  

By letting $\phi\in [-\frac{\pi}{2}, \frac{\pi}{2}]$ denote the steering angle and 
considering a Uniform Linear Array (ULA) with $d_e = \lambda/2$ inter-element antenna spacings, the Tx/Rx array response is given by  $\av(\phi)$, where $\av(\phi)=(a_1(\phi), \dots, a_{N_a}(\phi))^\transp\in \CC^{N_a}$ with
 \begin{align}\label{eq:ULA}
	a_n(\phi)&= e^{j(n-1)\pi \sin(\phi)},\;\; n=1,\dots, N_a .
\end{align}  
Since this paper focuses on the radar processing, we consider the channel model for the backscatter signal.  
For the case of $P$ targets, this is given by the superposition of $P$ rank-1 channel matrices, each of which corresponds to the \ac{LoS} propagation
from the Tx array to each target and back to the radar Rx array along the same \ac{LoS} path. This results in the 
$\Na\times\Na$ time-varying MIMO channel with matrix impulse response given by \cite{vitetta2013wireless}
\begin{align}\label{eq:Channel}
	\Hm (t, \tau) =  \sum_{p=0}^{P-1} \rho_p \av(\phi_p) \av^\herm (\phi_p)\delta(\tau-\tau_p)  e^{j2\pi \nu_p t}\,,
\end{align}
where for each target $p$,  $\rho_p$ is a complex channel gain including the \ac{LoS} pathloss and the radar cross-section coefficient,  $\nu_p = \frac{2 v_p f_c}{c}$ is the round-trip Doppler shift, 
$\tau_p=\frac{2r_p}{c}$ is the round-trip delay (time of flight), 
and $\phi_p$ denotes the \ac{AoA}.\footnote{In the expressions of $\nu_p$ and $\tau_p$, $c$ denotes the light speed, 
$v_p$ is the velocity component of the $p$-th target in the radial direction with respect to the radar receiver, and $r_p$ is the distance between the $p$-th target and the radar receiver.}
We assume that the channel parameters $\{\rho_p, \phi_p, \nu_p, \tau_p\}_{p=1}^P$ remain constant over the coherence processing interval of $B$ time-frequency blocks, where each time-frequency block consists of a frame of  (roughly) $W T_{\rm frame}$ signal dimensions, with $T_{\rm frame}$ denoting the frame 
duration. 

\subsection{OTFS Modulation}\label{subsec:OTFS-Input-Output}
%Motivated by \ac{mmWave} automotive applications, 
We consider the \ac{OTFS} modulation format as it is known to be robust to high Doppler shifts and efficient in the presence of sparse channels in the Doppler-delay domain \cite{gaudio2020effectiveness}. 
\ac{OTFS} is a multicarrier scheme with $M$ subcarriers with separation $\Delta f$, such that the total 
bandwidth is $W = M\Delta f$. We let $T$ denote the symbol time and $N$ denote the number of \ac{OTFS} symbols 
per frame, yielding a frame duration of $T_{\rm frame}=NT$. We also consider $T\Delta f = 1$, which is typical in most \ac{OTFS} literature \cite{gaudio2020effectiveness,hadani2017orthogonal,raviteja2018interference}. 
Let $\Ns$ denotes the number of data streams per frame to be sent by the \ac{BS}, 
where $\Ns = 1$ corresponds to the broadcasting of a single data stream (Discovery mode) 
and $1 < \Ns\leq \Nrf$ corresponds to the transmission of $\Ns$ individual users' data streams (Tracking mode).

In what follows, we derive the relation between the block of data symbols and the signal at the radar receiver. Since OTFS is a linear modulation and the propagation channel is a linear (time-varying) system, 
this relation will be a linear mapping.  
For simplicity of exposition,  we focus on a generic time-frequency block of $NM$ symbols and neglect the block index. Section \ref{sec:Joint-Detection-Param-Estimation} will consider the received signal 
across $B$ blocks explicitly. 

Following the standard derivation of the input-output relation of \ac{OTFS} (see, e.g., \cite{hadani2017orthogonal,gaudio2020effectiveness}),  
the $\Ns$-dimensional data symbol vectors 
$\{\xv_{k,l} \in \CC^{\Ns \times 1} : k\in[0:N-1], \; l\in[0:M-1]\}$, 
belonging to some suitable QAM constellation, are arranged in an $N \times M$ two-dimensional grid  
$\Gamma = \left\{\left(\frac{k}{NT},\frac{l}{M\Delta f}\right), k\in[0:N-1], l\in [0:M-1] \right\}$, referred to as the {\em Doppler-delay domain}. We can visualize $\left\{\xv_{k,l}\right \}$ as a $N \times M \times \Ns$ three-dimensional block of data, where each horizontal layer of dimensions $N \times M$ represents one data stream.  The Tx applies the \ac{ISFFT} layer by layer, converting
the Doppler-delay domain data block $\{\xv_{k,l}\}$ into the corresponding time-frequency data block $\{\Xm_{n, m}\}$, defined by 
\begin{equation}\label{eq:x-to-X}
\Xm_{n,m}=\sum_{k=0}^{N-1}\sum_{l=0}^{M-1}\xv_{k,l}e^{j2\pi\left(\frac{nk}{N}-\frac{ml}{M}\right)}, 
\end{equation}
for $n \in [0:N-1]$, $m \in[0:M-1]$. The symbols across time-frequency and data stream dimensions are uncorrelated, and we assume the average Tx power normalization 
\[ \EE[\Xm_{n,m} \Xm_{n,m}^\herm]= \frac{\Pav}{\Ns} \Id_{\Ns}, \;\;\; \forall \; (n, m). \]
Then, the Tx generates the $\Ns$-dimensional continuous-time signal
\begin{align}\label{eq:Tx-Signal-General}
\sv(t)  = \sum_{n=0}^{N-1} \sum_{m=0}^{M-1} \Xm_{n, m} g_{\rm tx}(t-nT) e^{j 2\pi m \Delta f(t-nT)}.
\end{align}
For \ac{mmWave} multiuser MIMO applications, different \ac{HDA} architectures have been considered in the literature
to handle the fact that the number $\Nrf$ of \ac{RF} chains  is generally significantly 
smaller than the number $\Na$ of antenna array elements
(e.g.,  \cite{HDA_Sohrabi,song2019fully}).  
Letting $\Fm\in \CC^{\Na \times \Ns}$ and $\Um\in \CC^{\Na \times \Nrf}$ denote the Tx and the Rx \ac{BF} matrices, respectively, from \eqref{eq:Tx-Signal-General} and \eqref{eq:Channel}, the $\Nrf$-dimensional continuous-time received signal at the radar Rx is obtained as\footnote{Here, we focus only on the useful part of the received signal expression. However, it is obvious that the received signal also contains  an additive white Gaussian noise term which is neglected here for the sake of brevity.}
\begin{align}\label{eq:Received-Signal-First}
	\rv(t) = \sum_{p=0}^{P-1} \rho_p {\Um}^\herm \av(\phi_p) \av^\herm(\phi_p) \Fm \sv(t - \tau_p) e^{j2\pi \nu_pt}\,.
\end{align}
The output of the Rx filter-bank adopting a generic receive shaping pulse $g_{\rm rx}(t)$ is given by
%\eqref{eq:y-received}, shown at the top of next page.
%\begin{figure*}
	\begin{align}\label{eq:y-received}
	\Ym(t,f) =& \int  \rv(t')g^*_{\rm rx} (t'-t) e^{-j2\pi ft'} dt' \nonumber\\
	=& \int_{t'} g^*_{\rm rx} (t'-t) \sum_{p=0}^{P-1} \rho_p {\Um}^\herm \av(\phi_p) \av^\herm(\phi_p)\Fm \sv(t'-\tau_p) e^{j2\pi \nu_pt'} e^{-j2\pi ft'} dt'\nonumber\\
	=& 
	%\sum_{n'=0}^{N-1}\sum_{m'=0}^{M-1} \sum_{p=0}^{P-1} 
	\sum_{p,n',m'} \rho_p {\Um}^\herm \av(\phi_p)\av^\herm(\phi_p) \Fm \Xm_{n', m'}\cdot \nonumber\\
	&\cdot\int_{t'} g^*_{\rm rx} (t'-t)  g_{\rm tx}(t'-\tau_p-n'T) e^{j 2\pi m' \Delta f(t'-\tau_p-n'T)} e^{j2\pi (\nu_p-f) t'} dt'
	\end{align} 
%\end{figure*}
By sampling at $t=nT$ and $f=m\Delta f$, we obtain
\begin{align}
\Ym_{n, m} &=\Ym(t, f)|_{t=n T}^{f=m\Delta f}=\sum_{n'=0}^{N-1} \sum_{m'=0}^{M-1}\Hm_{n,n',m,m'}\Xm_{n', m'}\,,  \label{ziocane}
\end{align}
where $\Hm_{n,n',m,m'}$ is given by 
%in \eqref{ziocane1}, shown at the top of next page.
%\begin{figure*}
%	\vspace{-0.5cm}
	\begin{align}
	\Hm_{n,n',m,m'} =& \sum_{p=0}^{P-1}  h_p {\Um}^\herm \av(\phi_p) \av^\herm(\phi_p)\Fm  
	\cdot\nonumber\\
	& \cdot C_{g_{\rm tx},g_{\rm rx}}((n-n')T -\tau_p, (m-m') \Delta f-\nu_p) e^{j 2\pi n' T \nu_p} e^{-j2\pi m \Delta f  \tau_p} \label{ziocane1}
	\end{align}
%\end{figure*}
and where we defined the cross ambiguity function $C_{u, v}(\tau, \nu) \eqdef \int_{-\infty}^{\infty} u(s) v^*(s-\tau) e^{-j 2\pi \nu s} ds$ as in \cite{matz2013time}, let $h_p \eqdef \rho_p e^{j 2\pi \nu_p \tau_p}$, and used the fact that $e^{-j2\pi mn'\Delta f T}=1$, $\forall n',m$, under the hypothesis $T\Delta f=1$.
Finally, the received signal in the Doppler-delay domain is obtained by the application of the \ac{SFFT}
\begin{align}
\yv_{k,l} = \frac{1}{NM} \sum_{n,m} \Ym_{n,m} e^{j2\pi\left(\frac{ml}{M}-\frac{nk}{N}\right)}.
\label{received-signal-expression}
%=\sum_{k', l'}\Gm_{k,k', l,l'}\xv_{k',l'},
\end{align}
%where the \ac{ISI} coefficient of the Doppler-delay pair $\left(k',l'\right)$ seen by sample %$\yv_{k,l}$ is given by
%\begin{equation}
%\Gm_{k,k',l,l'} =  \sum_p h_p \Psi_{k, k',l,l'}(\nu_p,\tau_p) {\Um}^\herm  %\av(\phi_p)\av^\herm(\phi_p) 
% \Fm \,,
%\end{equation}
In order to express (\ref{received-signal-expression}) in a more explicit and useful form, we 
defined the general Doppler-delay crosstalk coefficient
	\begin{align}\label{eq:Psi-Mat}
	\Psi_{k, k',l, l'}(\nu,\tau) \eqdef& \!\!\! \sum_{n, n', m, m'}\!\!\! \frac{C_{g_{\rm rx}, g_{\rm tx}}((n-n')T -\tau, (m-m') \Delta f-\nu)}{NM}  \cdot\nonumber\\
	& \cdot e^{j 2\pi n' T \nu} e^{-j2\pi m \Delta f  \tau}e^{j 2\pi \left(\frac{n'k'}{N}- \frac{m'l'}{M}\right)}e^{-j 2\pi\left(\frac{nk}{N}-\frac{ml}{M}\right)}
	\end{align}
%A simplified version of $\Psi_{k, k',l, l'}(\nu,\tau)$, obtained by
%letting ${g_{\rm tx}(t)$ and ${g_{\rm rx}(t)$ be rectangular pulses of length $T$ and assuming that $\max_p \{\tau_p\}< T$, can be found in \cite[eq.~(15)]{gaudio2020effectiveness}.
%\footnote{Under this assumption on pulse $g_{\rm rx}(t)$, the noise samples affecting the received samples %$\yv[k,l]$ are uncorrelated.}
Then, using \eqref{ziocane}, \eqref{ziocane1}, with  \eqref{eq:Psi-Mat}, in \eqref{received-signal-expression}, the the signal component of the 
channel output at a given Doppler-delay pair $k, l$ can be written as 
\begin{align}\label{ykl-explicit}
    \yv_{k,l} &=  \sum_{p=0}^{P-1} h_p {\Um}^\herm  \av(\phi_p)\av^\herm(\phi_p) \Fm  
    \sum_{k', l'} \Psi_{k, k',l,l'}(\nu_p,\tau_p) \xv_{k',l'}\,. 
%    &\approx \sum_{p=0}^{P-1} h_p {\Um}^\herm\av(\phi_p)\av^\herm(\phi_p)\fv_p \sum_{k', l'} \Psi_{k, k',l,l'}(\nu_p,\tau_p) x_{k',l'}[p]. 
\end{align}
In Discovery mode, the \ac{BS} sends a single data stream ($\Ns=1$) through a beamforming vector (i.e., $\Fm = [\fv]$ is formed by a single column). The beamforming vector $\fv$ 
is designed to uniformly cover a given wide angular sector as the \ac{BS} has no a priori knowledge of the location of the targets. 
In Tracking mode, the \ac{BS} sends $\Ns = P \geq 1$ data streams through a beamforming matrix $\Fm = [\fv_1, \dots, \fv_{\Ns}]$ where $\fv_q$ denotes the $q$-th column of $\Fm$ associated to the $q$-th data stream.  Appendix \ref{appendix:BF} presents the general method used in this work to design the Tx beamforming vectors $\fv$ and the columns of $\Um$.  
%
%At a given Doppler-delay pair $k, l$, the signal component of the 
%channel output is given by 
%\begin{align}\label{eq:381}
%    \yv_{k,l} &=  \sum_{p=0}^{P-1} h_p \Um^\herm \av(\phi_p)\av^\herm(\phi_p) \fv 
%    \sum_{k', l'} \Psi_{k, k',l,l'}(\nu_p,\tau_p) x_{k',l'}.
%\end{align}
%These \ac{BF} vectors are designed to generate narrow beams pointed in the directions of the %corresponding users, again with the method of Appendix \ref{appendix:BF}.  
%At a given Doppler-delay pair $k, l$, the signal component of the 
%channel output is given by 

We remark that \eqref{ykl-explicit} correspond to a single-input (for $\Ns=1$) or multiple input (for $\Ns = P > 1$) multiple-output channel with \ac{ISI}. where the \ac{ISI} occurs in the Doppler-delay domain.
%Also, we remark here that in Tracking mode the $P$ users served in spatial division multiple access are chosen %(by some multiuser MIMO scheduling/grouping scheme \cite{nam2014joint}) 
%to  to be sufficiently separated in the angle domain such that $\av^\herm(\phi_p)\fv_{q} \approx 0$ for $p\neq %q$ holds, such that the interference on user $p$ caused by the data stream sent to user $q \neq p$ is small.  %From the radar estimation viewpoint, this implies that in this scenario the (known) targets are always clearly %distinguishable in the angle domain.  

\section{Beam-space MIMO radar with multiple block reception}\label{sec:BeamDesign}

The fact that the number of RF chains $\Nrf$ is typically much smaller than 
the number of array antenna elements $\Na$ yields the following fundamental problem:
if the columns of the reduction matrix $\Um$ correspond to narrow beams with a high BF gain, the AoA domain is not well explored (e.g., some targets might be missed). 
In contrast, if the columns of $\Um$ correspond to wide angle beamforming patterns, 
the SNR at each receiver RF chain may be too low and the spatial resolution of each observation may be too coarse. 

In order to circumvent this problem,  we consider the joint processing of $B$ consecutive blocks, where the reduction matrix varies from one block to another (we let $\Um_b$ denote the 
reduction matrix in block $b\in [B]$).
The idea is that while each reduction matrix 
consists of a set of narrow beams, the ensemble of $B$ blocks is able to explore the \ac{FoV}
of interest without ``holes''. 
At each block $b$, we obtain an $\Nrf$-dimensional observation, 
where each dimension corresponds to a beam pattern defined by a column of $\Um_b$.
Hence, we refer to this approach as \textit{beam-space} MIMO radar. 
%The channel input-output relation in each block is described in \eqref{eq:381} and \eqref{eq:385} (depending on the considered mode), where the Tx beamforming matrix $\Fm$ is left invariant over the blocks, 
%while the sequence of reduction matrices is denoted by $\{\Um_b: b = 1, \ldots, B\}$.

\begin{figure}
	\centering
	\includegraphics[scale=0.89]{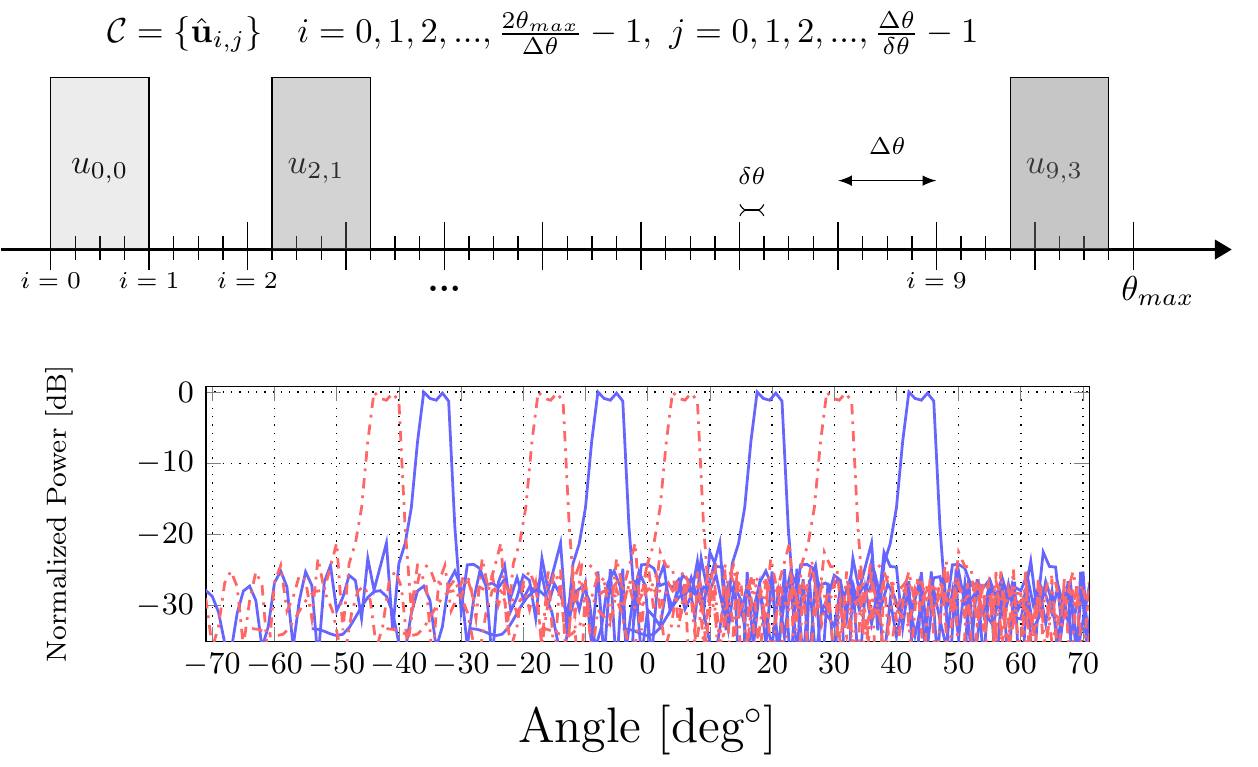}
	\caption{(Top) Illustrative description of the atoms of the BF codebook. Without loss of generality, we assume a symmetric \ac{FoV} covering $[-\theta_{\text{max}},\theta_{\text{max}}]$. (Bottom) An example with $\Nrf = 4$ beams per block for  $B=2$ consecutive blocks (corresponding to the different colors), and a total of $\Nrf B = 8$ explored directions.}
	\label{fig: strategy}
\end{figure}

%Given that no a priori information on the angular location of the targets is available during Discovery mode,\footnote{Beyond the fact that the targets of interest are in the \ac{FoV} of the transmit beam defined by $\fv$.} 
We consider a \ac{BF} codebook formed by a set of flat-top beams to span the required \ac{FoV}, designed to provide sufficiently large \ac{BF} gain (and therefore maintain a good receiver SNR) over a given angular span as compared to \textit{sharp} ``Fourier'' beams (i.e., beamforming vectors of the type of the ULA response vector defined in \eqref{eq:ULA}). 
Let $\Omega =[\theta_{\text{min}},\theta_{\text{max}}]$ denote the \ac{FoV} and, for given step $\Delta \theta$, we divide $\Omega$ into an integer number $\frac{\mid\Omega \mid}{ \Delta\theta}$ of 
intervals of size $\Delta \theta$. Each interval is further partitioned into an integer number 
$\frac{\Delta\theta}{\delta\theta}$ of subintervals of size $\delta \theta$. 
We let $\mathcal{C} \coloneqq \{ \widehat{\uv}_{i,j}~,~ i\in[0:\frac{\Omega}{\Delta\theta}-1],~j\in[0:\frac{\Delta\theta}{\delta\theta}-1]\}$ denote a \ac{BF} codebook, where each atom $\widehat{\uv}_{i,j} \in \CC^{\Na\times 1}$ is a 
direction-shifted version of the fundamental flat-top beam $\widehat{\uv}_{0,0}$ of width $\Delta\theta$ (designed using the method in Appendix \ref{appendix:BF}), with beam center direction 
given by $\theta_{\text{min}} + i\Delta\theta + j\delta\theta$. The parameters 
$\Delta\theta$ and $\delta\theta$ are selected to seek a good trade-off between BF gain, angle coverage, and complexity of the beamforming codebook.
%$\boldsymbol{\Sc}=\{\Sc_{i,b},~~i=1,..,\Nrf,~ b=1,...,B\}$
We have constructed pseudo-random sequences of $B$ reduction matrices such that 
at every block $b$, $\Um_b$ consists of ``non-overlapping'' atoms from $\mathcal{C}$, (i.e., such that any two $n \neq n'$ columns of $\Um_b$ satisfy $\uv_{b,n}^\herm \uv_{b,n'} \approx 0)$, and over the $B$ blocks the union of the covered angular span 
is maximal. An example of such \textit{multi-directional} beam patterns is illustrated in Fig.~\ref{fig: strategy}. 

%To formulate this problem, consider the set $\boldsymbol{\Sc}$ consisting of points corresponding to the center of the atoms selected from  $\mathcal{C}$. The aim is to partition subsets $\Sc_b \in \boldsymbol{\Sc}, ~ b=\{1,...,B\}$, each consisting of $\Nrf$ points. The points in each  subset $\Sc_{b}$ should be spread out as far as possible, i.e., we need to find points that have maximized minimum distances such that none of the beams overlap until the entirety of $\Omega$ is spanned. To this aim, we start with  a  random  subset  of $\Nrf$ points from $\boldsymbol{\Sc}$ and successively move each point to a better position which maximizes the distances to obtain $\Sc_{b}=(\Sc_{b,1},...,\Sc_{b,\Nrf})$. The minimum shift value of the points to increase the min-dist is restricted to $\delta \theta$. To maximally cover the \ac{FoV}, $\Sc_{b}$ are constructed sequentially by first optimizing $\Sc_{1}$ considering all points in $\mathcal{S}$, then $\Sc_{2}$, considering the points in $\boldsymbol{\Sc} \setminus \Sc_{1}$, and so on such that $\bigcup \Sc_{b}$ maximally covers $\Omega$. Eventually, columns of $\Um_b$ representing beamforming codewords are constructed as unit norm vectors $\Um_b = (\uv_{b,1},...,\uv_{b,\Nrf}) \in \CC^{\Na\times \Nrf}$ with columns given by 
%\begin{align}
%    %\uv_{n,b} = \frac{1}{\sqrt{\beta}}\sum_{i,j \in \Sc_{n,b}} \widehat{\uv}_{i,j}~, ~~~ %n=1,\ldots,\Nrf 
%    \uv_{b,n} = \widehat{\uv}_{i,j}~, ~~i,j \in \Sc_{b,n},~~ n=1,\ldots,\Nrf 
%\end{align}
%such that $\| \uv_{b,n} \| = 1$.

Extensive system simulations show that the system performance strongly depends  on $B$ for given $\Nrf$ and $\Na$ but, for relatively large $B$, it is almost independent on the specific pseudo-random choice of the matrices $\{\Um_b\}$ constructed according to the above principle. 

%\begin{remark}
%The rationale behind the use of randomized  beam patterns is suggested by the analogy with %compressed sensing. As in a compressed sensing problem, we wish to find the position 
%of  targets that are sparse in the angle domain. 
%It is well-known that in order to estimate a sparse vector from a small number of linear %projections, the projection matrix must satisfy certain properties (e.g., the Restricted %Isometry Property(RIP) \cite{CANDES2008}). 
%Furthermore, it is also well-known that the explicit deterministic construction of good %sensing matrices is a very difficult and yet open problem, but there exist large classes of %random matrices achieving very good performance with high probability. Our problem is not %equivalent to standard compressed sensing because 1) we consider joint target detection and %parameter estimation, where the observation model is not a simple linear underdetermined 
%projection of a sparse vector; 2) we use a Maximum-Likelihood approach which is not formulated %as convex optimization problem as in standard compressed sensing. Nevertheless, the intuition %that appropriately designed randomized measurements in the angle domain yield good statistics %for the detection and estimation problem at hand is justified and validated by the results and %comparisons provided in Section \ref{sec:Numerical-Results}.
%\hfill $\lozenge$
%\end{remark}

\begin{remark}  \label{remark1}
An alternative to the use of directive beams in the \ac{HDA} set-up discussed above consists of directly sampling $\Nrf$ antennas per block. 
We refer to this as the {\em Antenna Selection} scheme, where 
the dictionary $\Cc$ is formed by the columns of an $\Na \times \Na$ identity matrix such that, at each block $b \in[B]$, each column of $\Um_b$ consists of zeros except a single one corresponding to the antenna port being sampled.  Another alternative for beam-space MIMO radar consists 
of using a grid of \ac{DFT} beams, i.e., the dictionary $\Cc$ is formed by the columns of a $\Na \times \Na$ unitary \ac{DFT} matrix. 
In Section \ref{sec:Cramer-Rao-Bound}, Fig. \ref{fig: CRLB_methods}, a comparison between these alternatives in terms of the achieved \ac{CRLB} for parameter estimation will be provided. We will show that antenna selection and the \ac{DFT} dictionaries achieve generally worse performance than the proposed one  for two opposite reasons: antenna selection provides good angle exploration but very low SNR at each sampled antenna port. The \ac{DFT} dictionary provides very high \ac{BF} gains but too limited angular support. 
\hfill $\lozenge$
\end{remark}

\section{Joint Detection and Parameters Estimation}\label{sec:Joint-Detection-Param-Estimation}

We denote the true value of the parameters as $\mathring{\thetav} = \{\mathring{h}_{p}, \mathring{\nu}_{p}, \mathring{\tau}_{p}, \mathring{\phi}_{p}\}$ and use 
$\thetav = \{h_p, \nu_p, \tau_p, \phi_p \}$ to denote the arguments of the likelihood function.
%As said before, assuming that $\mathring{\thetav}$ is constant over $B$ OTFS blocks, target detection and parameter estimation are performed from the $B$-block observation.  
We shall write the received signal expression (\ref{ykl-explicit}) in a compact form by blocking the $NM$ Doppler-delay signal components into $NM \times 1$ vectors. In order to avoid notation ambiguity, we use underline to denoted blocked quantities. For each $b \in[B]$, we define the 
effective channel matrix of dimension ${\Nrf} NM\times \Ns NM$ as 
\begin{align}\label{eq:G-Matrix}
	\underline{\Gm}_{b}(\nu, \tau,\phi)\triangleq\left({\Um}^\H_b\av\left(\phi\right)\av^\herm(\phi)\Fm \right)\otimes\Psim (\nu,\tau)\,,
\end{align} 
where $\Psim(\nu,\tau)$ is defined such that $[\Psim(\nu,\tau)]_{kM+l, k’M+l'}= 
\Psi_{k, k’,l, l’}(\nu,\tau)$ for $k, k’\in [0: N-1]$ and $l,l'\in [0: M-1]$, 
where $\Psi_{k,k',l,l'}(\nu,\tau)$ is defined in \eqref{eq:Psi-Mat}, and
$\otimes$ is the Kronecker product. In \eqref{eq:G-Matrix}, $\Fm$ is formed by a single 
column (Discovery mode with $\Ns = 1$) or multiple columns (Tracking mode with $\Ns = P$).  
Thus, by stacking the $N\times M \times \Ns$ OTFS symbol block 
into a $\Ns NM$-dimensional vector $\underline{\xv}_b$ and defining the blocked 
output vector $\underline{\yv}_b$ of dimension ${\Nrf} NM \times 1$, the received signal takes on the form 
\begin{align}\label{eq:Received-signal}
\underline{\yv}_b = \left ( \sum_{p=0}^{P-1} \mathring{h}_{p} \underline{\Gm}_{b}(\mathring{\nu}_{p}, \mathring{\tau}_{p},\mathring{\phi}_{p}) \right ) \underline{\xv}_b  + \underline{\wv}_b, \;\;\; b\in [B],
\end{align}
where $\underline{\wv}_b$ denotes the \ac{AWGN} vector with independent and identically distributed entries of zero mean and variance $\sigma_w^2$. 

%%%%%%%%%%%%%%%%%%%%%%%%%%%%%%%%%%%%%%%%%%%%%%%%%%%%%%%%%%%%%%%%%%%%%%%%%%%%%
\subsection{Target detection and parameter estimation in discovery mode}\label{disc_paramEst}

In Discovery mode, the unknown number $P$ of targets are simultaneously illuminated by a single wide \ac{FoV} beacon signal. In this case, a near target may ``mask'' the presence of a far target with similar \ac{AoA}. Hence, we propose to detect the targets sequentially and, after a target is detected and its parameters are estimated, we use a \ac{SIC} approach in order to cancel it from the received signal and proceed to the detection of the next target. The procedure stops when no more targets are detected.\footnote{Alternative stopping criteria can be considered. For example, 
one may set a limit on the maximum number of targets to be detected at each detection cycle, since this operation is repeated periodically with a certain duty cycle. 
This depends on the specific application. Note that, in the estimation mode, each acquired user is served via a dedicated RF chain, therefore a limit for the number of detections can be bound by the number of available RF chains.}

At each detection step of the above described procedure, we are in the presence of a binary hypothesis testing where hypotheses $\Hc_0$ and $\Hc_1$ correspond to absence or presence of the $p$-th target only. 
In fact, when detecting a target and estimating the relevant parameters, the targets already detected are assumed to be already canceled from the received signal whereas the contribution of the remaining targets is assumed to be an additional noise. 
The observation under the two hypotheses is given by 
\begin{align} \label{eq:Hypo_test}
    \underline{\yv}_b = \begin{cases}
    \underline{\wv}_b & \;\; b \in[ B] \;\; \text{under $\Hc_0$}\\
    \mathring{h}_p \underline{\Gm}_{b}( \mathring{\nu}_p,\mathring{\tau}_p,\mathring{\phi}_p) \underline{\xv}_b  + \underline{\wv}_b
    & \;\; b \in[B] \;\; \text{under $\Hc_1$}\,.
    \end{cases}
\end{align}
In the following,  we neglect the arguments in $\underline{\Gm}_{b}(\tau_p, \nu_p,\phi_p)$ to avoid excessive  clutter in the notation.
The log-likelihood ratio for the binary hypothesis testing problem, multiplied by $\sigma_w^2$ for convenience, is given by 
\begin{align}
    \ell(h_p,\nu_p,\tau_p,\phi_p) & = \sigma_w^2 \log \frac{\exp\left ( - \frac{1}{\sigma_w^2} \sum_{b=1}^B \left \| \underline{\yv}_b - h_p \underline{\Gm}_b \underline{\xv}_b \right \|^2 \right )}{\exp \left ( - \frac{1}{\sigma_w^2} \sum_{b=1}^B \| \underline{\yv}_b \|^2 \right )} \nonumber \\
    & = 2 \Re\left \{ \left ( \sum_{b=1}^B \underline{\yv}_b^\herm \underline{\Gm}_b \underline{\xv}_b \right ) h_p \right \} - |h_p|^2 \sum_{b=1}^B \| \underline{\Gm}_b \underline{\xv}_b \|^2 \label{log-likelihood2}
\end{align}
Target detection is formulated here as a standard Neyman-Pearson hypothesis testing problem \cite{VPoor}, for which the solution that maximizes the detection probability subject to a bound on the false-alarm probability is given by the Likelihood Ratio Test
\begin{equation}\label{eq:likelihood_hypo_test}
    \ell(h_p,\nu_p,\tau_p,\phi_p)  \underset{\Hc_0}{\overset{\Hc_1}{\gtrless}} T_r,
\end{equation}
where the threshold $T_r$ determines the tradeoff between detection and false-alarm probabilities. 
Since the true value of the parameters is unknown, we use the Generalized Likelihood Ratio Test
\begin{equation}
    \max_{h_p,\nu_p,\tau_p,\phi_p} \; \ell(h_p,\nu_p,\tau_p,\phi_p)  \underset{\Hc_0}{\overset{\Hc_1}{\gtrless}} T_r.
\end{equation}
The maximization of \eqref{log-likelihood2} with respect to $h_p$ for fixed $\tau_p, \nu_p, \phi_p$ is immediately obtained  as
\begin{equation} 
\widehat{h}_p = \frac{\left ( \sum_{b=1}^B \underline{\yv}_b^\herm \underline{\Gm}_b \underline{\xv}_b \right )^*}{\sum_{b=1}^B \|\underline{\Gm}_b \underline{\xv}_b\|^2}.
\label{opth}
\end{equation}
Replacing \eqref{opth} into \eqref{log-likelihood2} we obtain the log-likelihood ratio in the form
\begin{equation}
    \ell(\widehat{h}_p,\nu_p,\tau_p,\phi_p) = \frac{\left | \sum_{b=1}^B \underline{\yv}_b^\herm \underline{\Gm}_b \underline{\xv}_b \right |^2}{\sum_{b=1}^B \| \underline{\Gm}_b \underline{\xv}_b \|^2}. \label{log-likelihood2S}
\end{equation}
For future use, we define the function $S(\nu,\tau,\phi)$ given by \eqref{log-likelihood2S}
after replacing $\nu_p \leftarrow \nu,\tau_p \leftarrow \tau,\phi_p \leftarrow \phi$. 
The proposed successive target detection, parameter estimation, and target signal cancellation works as follows.  We define the coarse Doppler-delay-angle search grid
$\Gamma \times \widehat{\Omega}$ where $\Gamma$ is the Doppler-delay grid defined in \ref{subsec:OTFS-Input-Output} and $\widehat{\Omega}$ is a suitably defined grid of discrete angles in the designed \ac{FoV} $\Omega$. The list of detected targets is initialized as ``empty''. For each detection step $p = 0,1,2,\ldots$, the algorithm repeats the following steps:
\begin{enumerate}
\item  Compute the adaptive threshold function $T_r(\nu,\tau,\phi)$  for all grid points
$(\nu,\tau,\phi) \in \Gamma \times \widehat{\Omega}$ according to the
Constant False Alarm Rate Detection (CFAR) approach. In particular, here we use 
the \ac{OS-CFAR} method, which is known to provide good performance in a realistic scenario when the statistic of noise and interference is not uniformly distributed across the three-dimensional grid (see e.g. \cite[Chapter 6.5]{richards2014fundamentals}).
The details of the computation of $T_r(\nu,\tau,\phi)$ are given in Appendix \ref{CFAR}.
\item Compare $S(\nu,\tau,\phi)$ with the threshold function and define the set of ``above threshold'' grid points 
\begin{equation} 
\Tc = \{ (\nu, \tau, \phi) \in \Gamma \times \widehat{\Omega} :  S(\nu,\tau,\phi) \geq T_r(\nu,\tau,\phi)\}. \label{set-T}
\end{equation}
\item If $\Tc = \emptyset$ (if $\Tc$ is empty, i.e., $S(\nu,\tau,\phi) < T_r(\nu,\tau,\phi)$ for all grid points), the algorithm exits. 
\item If $\Tc \neq \emptyset$ (and no other stopping criterion is reached), let
\begin{equation} 
(\widehat{\nu}_p, \widehat{\tau}_p, \widehat{\phi}_p ) = \arg \max_{(\nu,\tau,\phi) \in \Tc}  \; S(\nu,\tau,\phi), 
\end{equation}
and declare the new detected $p$-th target with coarse estimated parameters $(\widehat{\nu}_p, \widehat{\tau}_p, \widehat{\phi}_p)$. 
\item Refine the coarse estimate of the parameters over fine grid search localized in the neighborhood of 
$(\widehat{\nu}_p, \widehat{\tau}_p, \widehat{\phi}_p)$ in the 3-dimensional search space, and let 
$(\check{\nu}_p, \check{\tau}_p, \check{\phi}_p)$ denote the arg-max of $S(\nu,\tau,\phi)$ on the local search fine grid.
\item Replace $(\check{\nu}_p, \check{\tau}_p, \check{\phi}_p)$ into 
(\ref{opth}) and find the corresponding estimate $\check{h}_p$ of the channel coefficient. Then, subtract the $p$-th path signal contribution from the received signal, i.e., 
\[ \underline{\yv}_b \leftarrow \underline{\yv}_b - \check{h}_p \underline{\Gm}_b(\check{\nu}_p, \check{\tau}_p, \check{\phi}_p) \underline{\xv}_b, \;\;\; \mbox{for}\; b \in[B]. \]
Go back to Step 1 and repeat. 
\end{enumerate}

\subsection{Refined parameter estimation in tracking mode}  \label{sec:new-para-estimation}

In Tracking mode, the $P$ users served in spatial division multiple access are chosen 
by some multiuser MIMO scheduling/grouping scheme (e.g., see \cite{nam2014joint}) so that 
they are sufficiently separated in the angle domain thus suffering from very small
inter-user interference. 
It follows that, by design, we have $\av^\herm(\phi_p)\fv_{q} \approx 0$ 
for $p\neq q$. From the radar estimation viewpoint, this implies that in this scenario the (known) targets are always clearly distinguishable in the angle domain. Notice that there is no loss of generality in this assumption {\em precisely} because we are considering the tracking of already connected users, which are scheduled for data transmission and hence chosen (by the multiuser scheduler) to be separable in the angle domain. In other words, if two users are not separable in the angle domain, the \ac{BS} schedules them in different data frames. 

This implies that the matrix $\underline{\Gm}_b(\mathring{\nu}_p,\mathring{\tau}_p,\mathring{\theta}_p)$ defined in \eqref{eq:G-Matrix} can be partitioned into $P$ vertical slices of dimension $\Nrf NM \times NM$, 
where all but the $p$-th slice are $\approx 0$. We define the $p$-th vertical slice of the channel matrix as
\begin{equation} \label{Gp}
\underline{\Gm}_{b,p}(\mathring{\nu}_p,\mathring{\tau}_p,\mathring{\phi}_p)  = 
\left ( \Um_b^\herm \av(\mathring{\phi}_p) \av^\herm(\mathring{\phi}_p) \fv_p \right ) \otimes \Psim(\mathring{\nu}_p,\mathring{\tau}_p). 
\end{equation} 
The received signal (\ref{eq:Received-signal}) after neglecting the effect of the almost zero ``slices'' can be written as
\begin{equation} \label{eq:Received-signal-simp}
\underline{\yv}_b \approx 
\sum_{p=0}^{P-1} \mathring{h}_{p} \underline{\Gm}_{b,p}(\mathring{\tau}_{p}, \mathring{\nu}_{p},\mathring{\phi}_{p}) \underline{\xv}_{b,p}  + \underline{\wv}_b, \;\;\; b\in[B],
\end{equation}
where $\underline{\xv}_{b,p}$ is the $NM \times 1$ $b$-th symbol block of the $p$-th user data stream. 
We shall develop our \ac{ML}-based parameter estimation scheme under the assumption that \eqref{eq:Received-signal-simp} holds with equality. Of course, in simulation, we shall test the scheme 
with the true channel model given by \eqref{eq:Received-signal}. The excellent performance of the resulting estimator (closely approaching the \ac{CRLB}) demonstrates the validity of this approximation, which in turns yields a greatly simplified and low complexity estimation scheme. 

As before, we neglect the arguments in $\underline{\Gm}_{b,p}(\tau_{p}, \nu_{p}, \phi_{p})$. The log-likelihood function, neglecting irrelevant terms, is given by 
\begin{align}
    \Lambda(& \{h_p,\nu_p,\tau_p,\phi_p\})  = - \sum_{b=1}^B \left \| \underline{\yv}_b - \sum_{p=0}^{P-1} \underline{\Gm}_{b,p} \underline{\xv}_{b,p} \right \|^2 = \nonumber \\ &  - \sum_{b=1}^B \|\underline{\yv}_b \|^2 + 2 \Re \left \{ \sum_{p=0}^{P-1}  h_p^* \left ( \sum_{b=1}^B \underline{\xv}_{b,p}^\herm \underline{\Gm}_{b,p}^\herm \underline{\yv}_b \right ) \right \} - 
    \sum_{p=0}^{P-1} \sum_{q=0}^{P-1} h_p^* h_q \left ( \sum_{b=1}^B 
    \underline{\xv}_{b,p}^\herm \underline{\Gm}_{b,p}^\herm \underline{\Gm}_{b,q} \underline{\xv}_{b,q} \right ) \label{log_lkhd_main}
\end{align}
Defining the $P \times 1$ vector of path coefficients $\hv = (h_0, \ldots, h_{P-1})^\transp$, the vector of signal correlations $\rv$ with $p$-th element
\begin{equation} 
r_p = \sum_{b=1}^B \underline{\xv}_{b,p}^\herm \underline{\Gm}_{b,p}^\herm \underline{\yv}_b, 
\label{rr}
\end{equation} 
the $P \times P$ matrix $\Am$ with $(p,q)$ element
\begin{equation} 
A_{p,q} = \sum_{b=1}^B 
\underline{\xv}_{b,p}^\herm \underline{\Gm}_{b,p}^\herm \underline{\Gm}_{b,q} \underline{\xv}_{b,q}, 
\label{AA}
\end{equation} 
and neglecting the irrelevant first term in the RHS of \eqref{log_lkhd_main}, with some abuse of notation, the equivalent log-likelihood function can be written as
\begin{equation} 
\Lambda(\{h_p,\nu_p,\tau_p,\phi_p\}) = 2\Re\{ \hv^\herm \rv \} - \hv^\herm \Am \hv. 
\label{log_lkhd_1}
\end{equation} 
The maximization with respect to $\hv$ is readily obtained as
$\widehat{\hv} = \Am^{-1} \rv$.
Replacing this into \eqref{log_lkhd_1}, the reduced log-likelihood function with respect to the parameters of interest $\{\nu_p,\tau_p,\phi_p\}$ is given by the quadratic form
\begin{equation}
   \Lambda_1(\{\nu_p,\tau_p,\phi_p\}) = \rv^\herm \Am^{-1} \rv. 
   \label{log_lkhd_2}
\end{equation}
Notice that \eqref{log_lkhd_2} must be maximized with respect to the 
$3P$ parameter variables in order to find the \ac{ML} parameter estimator. In Tracking mode, the \ac{BS} has already a coarse knowledge of the parameters of each target (user) since it is transmitting data to them. Therefore, it knows
(with some coarse approximation) the \ac{AoA} (necessary to point the transmit beams), while the delay and Doppler shifts can be obtained (for example) from the data in the uplink. Nevertheless, even with a coarse knowledge of the parameters, a brute-force maximization of  \eqref{log_lkhd_2} is not feasible. For example, a search over a fine grid with 10 points per parameter 
around their coarse estimates yields already $10^{3P}$ evaluations of 
\eqref{log_lkhd_2}. For $P = 3$ this yields 1 billion of points! 

This problem is overcome here by noticing a further simplification of the likelihood function. 
Since the data blocks $\underline{\xv}_{b,p}$ are formed by independent zero mean random variables and the block size $NM$ is large, the signal correlation terms $A_{p,q}$ are negligible for $p \neq q$. Neglecting the off-diagonal terms in the matrix $\Am$ the reduced log-likelihood function becomes separable in the individual targets parameters. In fact, it is easily seen that under this simplification we obtain
\begin{equation}  
\Lambda_1(\{\nu_p,\tau_p,\phi_p\}) = \sum_{p=0}^{P-1} 
\frac{\left | \sum_{b=1}^B \underline{\yv}_b^\herm \underline{\Gm}_{b,p} \underline{\xv}_{b,p} \right |^2}{\sum_{b=1}^B \| \underline{\Gm}_{b,p} \underline{\xv}_{b,p} \|^2}.
\label{log_lkhd_3}
\end{equation} 
Each term in the sum in \eqref{log_lkhd_3} has a form similar to the function $S(\nu,\tau,\phi)$ defined in 
\eqref{log-likelihood2S} and can be maximized individually with respect to the corresponding parameters $\{\nu_p,\tau_p,\phi_p\}$ using the same 3-dimensional grid search as done for the target detection scheme. The numerical results in our simulations are based on this simplified ML-based scheme. 

%{\bf Approach 2:}
%We follow an iterative component-wise maximization approach, where we maximize $\Lambda_1(\{\nu_p,\tau_p,\phi_p\})$ with respect to the parameters of one target at a time, for fixed values of the other targets' parameters. Starting from a coarse initial estimate $\{\widehat{\nu}_p,\widehat{\tau}_p,\widehat{\phi}_p\}$
%(available in tracking mode as discussed before), at each iteration $i$, letting $p = i \mod P$, we update the $p$-th target parameters by maximizing with respect to $\nu_p,\tau_p,\phi_p$ the function
%\begin{equation} 
%\widetilde{S}(\nu_p,\tau_p,\phi_p) = |r_p|^2 [\Am^{-1}]_{p,p} + 2 \Re\left\{ 
%r_p^* \sum_{q \neq p} [\Am^{-1}]_{p,q} r_q \right\} 
%\label{S_signal}
%\end{equation} 
%where $[\Am^{-1}]_{p,q}$ denotes the $(p,q)$ element of $\Am^{-1}$ and the dependency on the %parameters 
%is implicit through the definition of $\rv$ and $\Am$ in \eqref{rr} and \eqref{AA}, respectively. Notice also that if $\Am$ is exactly diagonal, then the first term in the RHS of \eqref{S_signal} coincides with the $p$-th term in the sum in \eqref{log_lkhd_3}, and the second term disappear, i.e., in this case the two approaches coincide. 
%\input{MatixA_enteries}

%%%%%%%%%%%%%%%%%%%%%%%%%%%%%%%%%%%%%%%%%%%%%%%%%%%%%%%%%
\subsection{Cram\'er-Rao Lower Bound (CRLB)}\label{sec:Cramer-Rao-Bound}

We consider the \ac{CRLB} as a theoretical benchmark, in particular to evaluate the ``goodness'' of various alternative reduction matrix design (see Remark \ref{remark1}). We consider the case of a single target ($P = \Ns = 1$) and drop the index $p$ for simplicity of notation. 
Letting $A = |h|$ and $\psi = \angle(h)$ denote the amplitude and the phase of $h$, respectively, five real parameters, denoted by $\thetav =(A, \psi, \tau, \nu, \phi)$, shall be estimated. 
Let $\sv_{b, k, l}(\thetav)$ denote the noise-free received signal at Doppler-delay bin $(k,l)$ and block $b$, obtained by letting $\Um \leftarrow \Um_b$ and $P = 1$ in 
(\ref{ykl-explicit}). Since the signal is observed in \ac{AWGN}, we can use the general expression in \cite[Sec. 3.9]{kay1993fundamentals} to obtain the $5\times 5$ Fisher information matrix with $(i,j)$-th element given by 
\begin{align}\label{eq:Fisher}
[\Id(\mathring{\thetav}) ]_{i,j} =
%\frac{2}{\sigma_w^2} \Re\left\{ \sum_{b=1}^B\sum_{t=1}^{\Nrf} \sum_{n=0}^{N-1}\sum_{m=0}^{M-1} \left[\frac{\partial s_{b,p}[n ,m, %t]}{\partial \theta_i}\right]^* \left[\frac{\partial s_{b,q}[n, m,t]}{\partial \theta_j}\right]\right\}\,, 
 \Re\left\{ \EE \left [ \sum_{b=1}^B \sum_{k=0}^{N-1}\sum_{l=0}^{M-1} \left ( \frac{\partial \sv_{b, k, l}(\thetav)}{\partial \theta_i}\right )^\H  \left (\frac{\partial \sv_{b, k, l}(\thetav)}{\partial \theta_j}\right ) \right] \right\}
 \Bigg|_{\thetav=\mathring{\thetav}} , 
\end{align}
% \begin{align} \label{eq:Fisher_elmn}
% s_{b,p}[n,m,t]&= A_p e^{j \psi_p} \uv_{b, t}^\H \av(\phi_p)\av^\H (\phi_p)\fv \sum_{k=0}^{N-1}\sum_{l=0}^{M-1}\Psi_{n, k, m, %l}(\tau_p, \nu_p) x^{(b)}_{k, l}\,,
% \end{align}
%where $\uv_{b, t}$ denotes the $t$-th column of $\Um_b$, $\Psi_{n, k, m, l}(\nu,\tau)$ 
%is defined in \eqref{eq:Psi-Mat}, and $x^{(b)}_{k,l}$ is the $(k,l)$-symbol of the data block $\xv_b$. 
The complete derivation of the Fisher Information matrix requires straightforward but very cumbersome algebra and is presented in detail in Appendix \ref{app:CRLB}. 
The desired \ac{CRLB} is then obtained by taking the diagonal elements of the inverse Fisher information matrix.

\begin{figure}[h]
	\centering
	\includegraphics[scale=1.0]{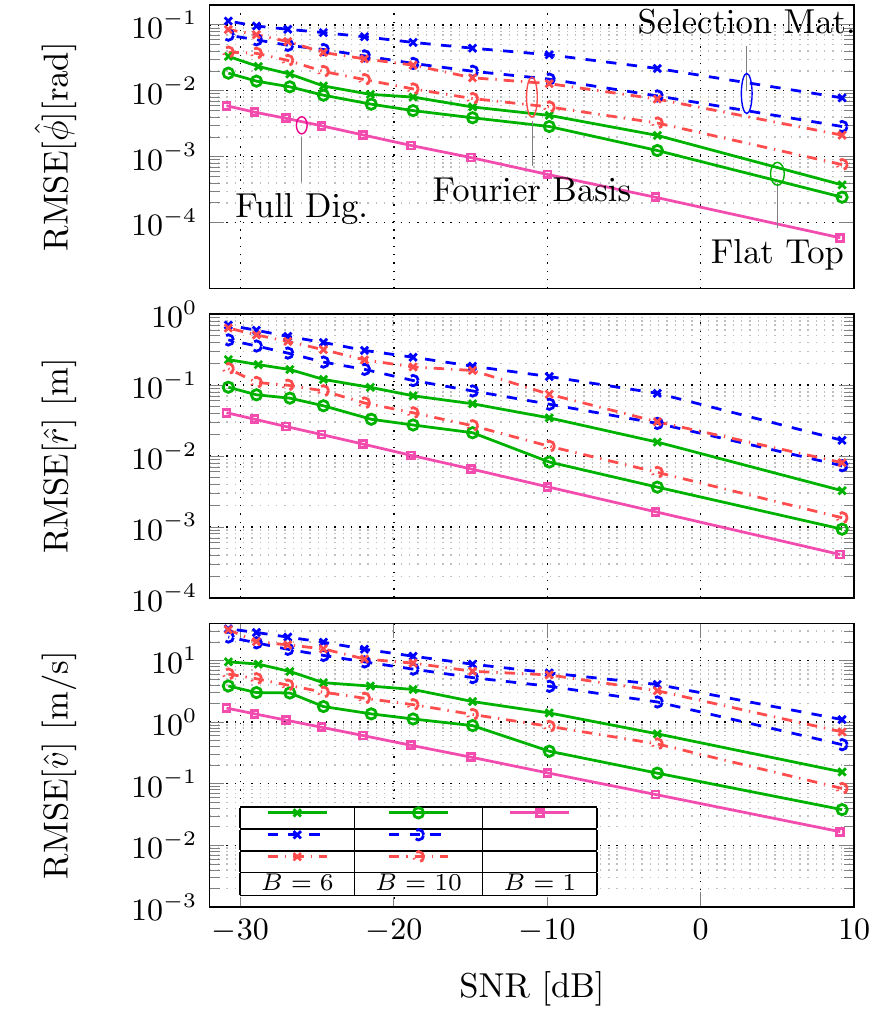}
	\caption{Comparison of the receive-beamforming strategies over \ac{SNR} and varying number of integration blocks. \ac{SNR} values are defined according to \eqref{eq:SNR-BBF-Formula}. }
	\label{fig: CRLB_methods}
\end{figure}

 In Fig.~\ref{fig: CRLB_methods} a comparison of the \ac{CRLB} for different alternative 
 designs of the reduction matrices $\{\Um_b\}$  and for the system parameters defined in Table~\ref{tab:System-Parameters} is provided. 
 This comparison considers three approaches: 1) Proposed method with \textit{Flat-Top} beams. 2) Random selection of $\Nrf$ antenna elements at each block. 
 3) A strategy similar to the proposed one, with beams from a 
 Fourier dictionary (DFT grid of beams of size $\Na$). 
 Additionally, as a reference, a fully digital system with $\Nrf = \Na$ and $\Um = \Id_{\Na}$ with only a single integration block is considered. Although this is highly impractical for implementation (as pointed out in Section \ref{sec:Introduction}), it is provided here as a useful term of comparison.
 %As a further remark, since the choice of $M,N$ and $\Delta f$ affect the parameter estimation capability, the \acro{CRLB} can be used to indicate the choice of these parameters to meet a specific system performance requirement. 

%It's important to note that when the considered FoV is wide, even though the Fourier basis beams provide very high beamforming gain around the main lobes, the generated pattern constitutes many \textit{holes} where an adequate beamforming gain is not available.

%Beyond the comments already made in Remark \ref{remark1}, here we provide further insight
%into the selection of parameters defining the proposed \ac{BF} codebook. 
%As detailed in section \ref{sec:BeamDesign}, $\Delta\theta$ and $\delta \theta$ determine the %tradeoff between beam-space exploration (which relates directly to latency, i.e. number of blocks %required to adequately cover the \ac{FoV}) and the \ac{BF} gain.
%$\Delta\theta$ determines the beam width of each dictionary element (thus, the\ac{BF} gain) 
%and $\delta \theta$ affects the sharpness of the function $S(\nu,\tau,\phi)$ around its true value %
%$(\mathring{\nu},\mathring{\tau},\mathring{\phi})$.
%This is caused by the fact that left and right shifts of the beam that encompasses the target can %lead to a sharpening effect on the beam over multiple integration blocks. %Fig.~\ref{fig:Shift-params} depicts these effects.
%
%\begin{figure}
%	\begin{subfigure}{.6\columnwidth}
%		\centering
%		\includegraphics[scale=1]{Figures/Pd_vs_shifts.pdf}
%		\caption{P$_{\rm d}$ comparison for values of $\Delta\theta , \delta\theta$ and as function of \ac{SNR}.}
%		\label{fig:Shift-params-a}
%	\end{subfigure}
%	\begin{subfigure}{.3\columnwidth}
%		\centering
%		%\vspace*{-0.55cm}
%		\includegraphics[scale=0.85]{Figures/PeakWidth_Beta.pdf}
%		\caption{Effect of minor shift on the main peak width.}
%	    \label{fig:Shift-params-b}
%		\end{subfigure}
%		\caption{System parameters: $\Delta\theta , \delta\theta$}
%		\label{fig:Shift-params}
%\end{figure}

%%%%%%%%%%%%%%%%%%%%%%%%%%%%%%%%%%%%%%%%%%%%%%%%%%%%%%%%%%%%%%%%%%%%%%%%%
\section{Numerical Results}\label{sec:Numerical-Results}
\renewcommand{\arraystretch}{1.3}

\begin{table}[ht]
	\caption{System parameters}
	\centering
	\begin{tabular}{|c|c|}
		\hline
		$N=64$ & $M=64$ \\ \hline
		$f_c=28.25$ [GHz] & $W=64$ [MHz] \\ \hline
%		$v_{\mathrm{res}}\simeq 95$ [m/s] & $r_{\mathrm{res}}\simeq 2.3 $ [m]\\ \hline
%		$v_{\mathrm{max}}=N\cdot v_{\mathrm{res}}$ & $r_{\mathrm{max}}=M\cdot r_{\mathrm{res}}$ \\ \hline
		$\Pav=24$ [dBm] & $\sigma_{\mathrm{rcs}}=1$ [m$^2$] \\ \hline
		Noise Figure (NF) $=3$ [dB] & Noise PSD $N_{0}$ = $2\cdot10^{-21}$ [W/Hz] \\ \hline
		$\Na=64$ & $\Nrf=4$ \\ \hline
	\end{tabular}
	\label{tab:System-Parameters}
\end{table}

\subsection{Simulation Setup}

We set the number of \ac{RF} chains to $\Nrf=4$, such that a single equipment (e.g., \ac{BS}) is able to jointly track and communicate to $\Nrf$ distinct targets (or groups of targets), while $\Nrf\ll\Na$. A summary of the system parameters is provided in Table \ref{tab:System-Parameters}. %The following system assumptions are in order:
%\begin{itemize}
%	\item Given the choice of a \ac{mmWave} communication, we assume a single \ac{LoS} path between the Tx and the radar target \cite{kumari2018ieee,nguyen2017delay,grossi2018opportunisticRadar} as explained in Section \ref{subsection:PhysicalModel}.
%	%This is motivated by the fact that any possible scattering component different from the %\ac{LoS} generally brings much lower power, given by additional reflections of the echo %signal.
%	\item Any backscattered power to the radar Rx is considered a possible target. The objective is to sense the surrounding environment, and the differentiation between active targets and obstacles is a post-processing decision. Clearly, in Tracking mode, communication is established only towards active targets. 
%	\item We consider the complete blockage of signal propagation to the first object hit. This assumption is completely fulfilled in \ac{mmWave} communication scenarios.\footnote{Note that the proposed algorithm could be able to correctly distinguish more targets sharing the same angular direction, if separated in at least one other domain (Doppler or delay) \cite{gaudio2020effectiveness}.} 
%\end{itemize}
%Note that the aforementioned assumption are shared by many works in literature (see, e.g., \cite{grossi2018opportunisticRadar} and references therein). 

The radar two-way pathloss is defined as \cite[Chapter 2]{richards2014fundamentals}
$\mathrm{PL}=\frac{(4\pi)^3r^{4}}{\lambda^2}$,
% such that the standard \ac{PL} (see, e.g., \cite[Chapter 2]{richards2014fundamentals}) formula is obtained for $\eta=2$, while, at \ac{mmWave} at $f_c=24.25$ GHz \cite{braun2014ofdm,sturm2011waveform}, by properly taking into account additional gases absorption effects \cite{rappaport2012propagation}, a more precise formulation of the \ac{PL} requires $\eta=2.5$. 
and the resulting \ac{SNR} at the radar receiver is given by 
\begin{equation}\label{eq:SNR-BBF-Formula}
\mathrm{SNR}=\frac{\lambda^2\sigma_{\mathrm{rcs}}}{\left(4\pi\right)^3r^4}\frac{\Pav}{\sigma_w^2}\,,
% \mathrm{SNR_{a}}=\frac{\lambda^2\sigma_{\mathrm{rcs}}G_\mathrm{Tx}G_\mathrm{Rx}}{\left(4\pi\right)^3r^4}\frac{\Pav}{\sigma_w^2}\,,
\end{equation}
where $\lambda=\frac{c}{f_c}$ is the wavelength, $c$ is the speed of light, $\sigma_{\mathrm{rcs}}$ is the radar cross-section of the target in $\mathrm{m}^2$, $r$ is the distance between Tx and Rx, and $\sigma_w^2$ is the variance of the \ac{AWGN} with \ac{PSD} of $N_{0}$ in W/Hz. We choose $\sigma_{\mathrm{rcs}}=1$ [m$^2$] as an indicative value, while different choices for specific cases may be found in literature \cite{suzuki2000measurement,kamel2017RCSmodeling}.

\begin{remark}
In the radar literature, it is customary to consider the resolution limits of each parameter individually. For example, the target (radial) velocity with respect to the radar receiver 
is given by $v = \frac{\nu c}{2f_c}$, and the range (distance between the target an the radar receiver) is given by $r = \frac{\tau c}{2}$. The corresponding velocity, range \cite{RD_res} and approximate angular \cite{Auto_radars} resolutions, expressed in terms of the system parameters of Table \ref{tab:System-Parameters}, are given by
\begin{equation}\label{eq:Radar-Resolution}
v_{\mathrm{res}}=\frac{cW}{2NMf_c}\ \mathrm{[m/s]}\,,\:\:\:r_{\mathrm{res}}=\frac{c}{2W}\ \mathrm{[m]}\,,\:\:\: \Theta_{\mathrm{res}} = 1.22\frac{\lambda}{L}\ \mathrm{[rad]},
\end{equation}
where for the \ac{ULA} described in section \ref{subsection:PhysicalModel}, the electrical antenna length $L$ is equal to $\frac{\Na \lambda}{2}$. It can be observed that the velocity and range resolutions are directly proportional to target illumination time (total frame duration) and RF bandwidth, respectively. The single-parameter resolution is (approximately) the 
minimum spacing such that two targets are distinguishable (i.e., identifiable) 
in the domain corresponding to the given parameter. At this point, two important observations are in order: 1) in Discovery mode, since the proposed scheme performs a search over the three-dimensional 
parameter space, two targets become indistinguishable if they are separated by less than the resolution limit in {\em all} three parameters, Doppler, delay, and AoA. For example, 
two targets may be seen under the same distance and radial velocity, but at different AoA, and yet our scheme with successive detection can detect them with high probability (see for example Fig.~\ref{fig:Detection-prob-MRS} marked with $^{\dagger}$). 2) The resolution limits have little to do with the accuracy (in terms of Mean-Square Error) with which the parameters of detected targets can be estimated in Tracking mode. In fact, our \ac{ML}-based parameter estimator operates a sort of super-resolution estimation on a much finer search grid, and when not limited by the discretization of the grid, it can approach very closely the \ac{CRLB} for parameter estimation.  \hfill $\lozenge$
\end{remark}

\subsection{Simulation Results}

We consider separately the detection performance in Discovery mode and the parameter estimation performance in Tracking mode,

\subsubsection{Discovery Mode}

In Fig.~\ref{fig:Detection-prob-MRS}, we illustrate the detection probability $P_{\rm d}$ of the proposed method as a function of target range and varying number $B$ of integration blocks for single and multi-target scenarios. 
Note that range and \ac{SNR} are related through \eqref{eq:SNR-BBF-Formula}.  
Given that the initial target acquisition has a direct impact on the latency with which new users can connect to the \ac{BS} (e.g., during a handover operation) we consider 
using only relatively small values of $B=\{6,10\}$. The plots are obtained by Monte Carlo simulations at each SNR point, where the angles and Doppler shifts of the 
targets are randomly changed within a wide angular \ac{FoV} $=[-45^{\circ},45^{\circ}]$ and radial velocity range $[10-60]$ m/s, respectively. 
We consider that a target is correctly detected if the estimated AoA, $\hat{\phi}_p$ 
fulfills $|\hat{\phi}_p- \phi_p| \leq \epsilon$, with $\epsilon = 0.5^{\circ}$, since this simulation corresponds to the coarse estimation stage in Discovery mode. 
We can observe the effect of integration gain on the target detection performance, for which the performance improves with the number of blocks $B$.

\begin{figure}[h]
	\centering
	\includegraphics[scale=1.0]{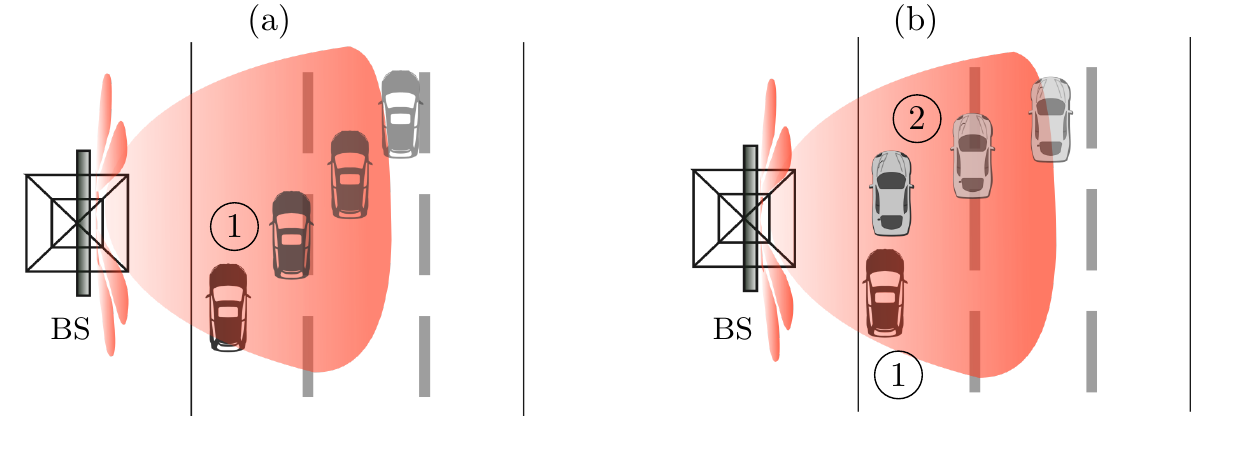}
	\hspace{-1cm}
	\caption{(a) Movement of a single target inside the considered \ac{FoV}. (b) Movement scenario of two targets inside the considered \ac{FoV}, where one target is located at a fixed closer distance and the second changes its position.}
	\label{fig:Detection-animation}
\end{figure}

\begin{figure}[h]
	\centering
	\includegraphics[scale=0.9]{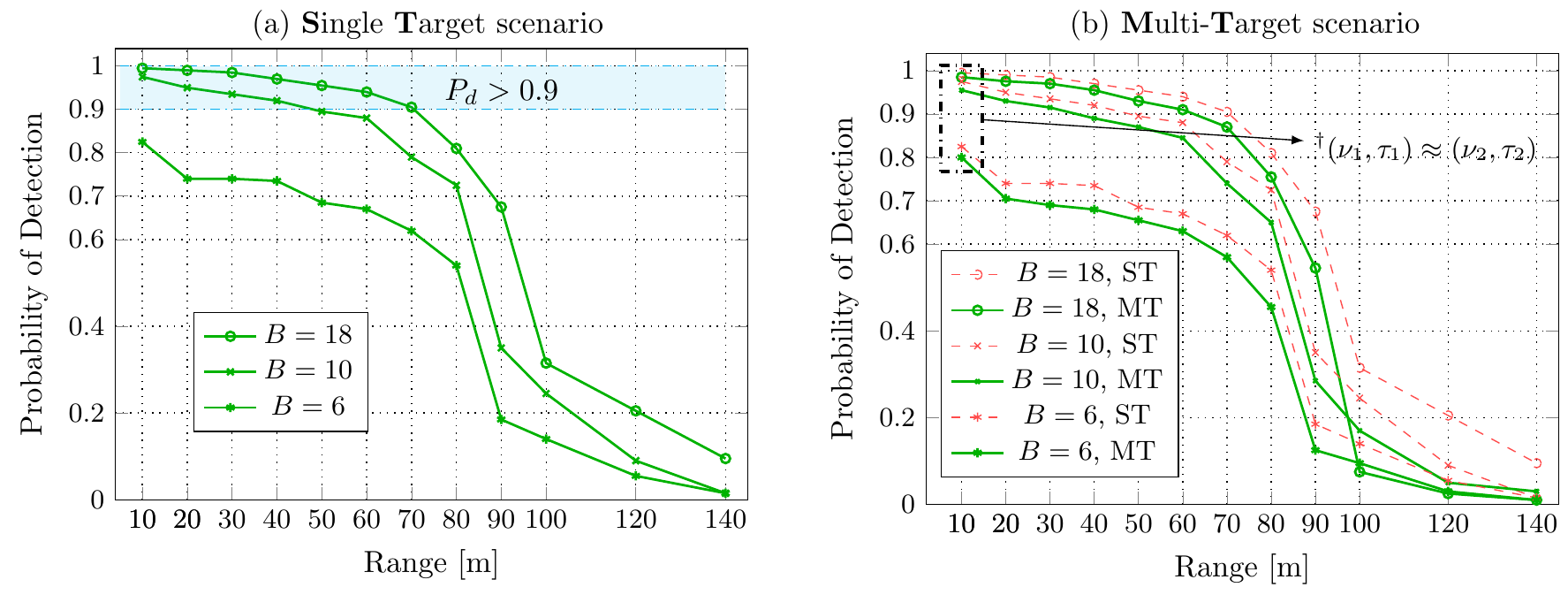}
	\caption{Probability of detection of targets Vs. range within an illuminated angular \ac{FoV} of $90^{\circ}$, as depicted in fig.~\ref{fig:Detection-animation}, with varying $B$. Plot (b) depicts a two target scenario where $P_d$ for the second target after detection and removal of the first target is reported.}
	\label{fig:Detection-prob-MRS}
\end{figure}

%  {\RED [PLEASE LABEL THE TWO PLOTS AS a) and b) and make them on the same line, with taller and thinner figures.
% 	Have the cartoonish representation of the 1-target and 2-target topologies on a different figure, again with a) and b)]}
% \begin{figure}
% 	\centering
% 	\includegraphics[scale=0.65]{Figures/peak_vs-Thr_90m_B17_F2and3.pdf}
% 	\caption{Comparison of signal and threshold for a single target simulated at 90m at with $B=17$ and  different number of fingers per RF chain.}
% 	\label{fig:peakVsThreshold}
% \end{figure}

% First, note that, by considering an angular coverage of $90^\mathring$ degrees (blue line), the maximum range to correctly identify a target, limited by the pathloss and thus different from the theoretical limit indicated in Table \ref{tab:System-Parameters}, is about $110$ m. 

% \begin{figure}[h]
% 	\centering
% 	\includegraphics[width=12cm]{Journal_Discovery_plus_SIC.pdf}
% 	\caption{The simulated scenario in Discovery mode with a very wide \ac{FoV}. (a) Single target (b) Multiple target scenario with SIC algorithm.}
% 	\label{fig:schematic_sim_scnes}
% \end{figure}

 To demonstrate the effectiveness of the \ac{SIC} technique of 
 Section \ref{disc_paramEst}, both a one-target and a two-target scenarios are considered.
 In the two-target case, one target is located at a close distance of $ r = 10$ m, and the second target is located at a distance varying from $r = 10$ m to $r = 140$ m. 
 The closer target creates a masking effect of the second target.
By comparing the detection probability $P_d$ versus range in the case of a single target in Fig.~\ref{fig:Detection-prob-MRS} (a) with that of the second target 
in the two-target case in Fig.~\ref{fig:Detection-prob-MRS} (b), 
plotted versus the range of the second target while the first is at fixed short distance, 
we notice that the proposed SIC scheme is able to cope well with the masked target effect. 
In fact, $P_d$ for the second target in Fig.~\ref{fig:Detection-prob-MRS} (b) is very close to $P_d$ for the single target case in Fig.~\ref{fig:Detection-prob-MRS} (a), showing that the presence of a close masking target with strong near-far effect incurs only a small degradation, at least in the relevant range up to 80 m. It should be noted that the effective reliable detection range is a function of transmit power, which can be extended by increasing the \ac{Tx} power.     
%, which can be removed using the SIC technique. $P^{(2)}_{\rm d}$ for the second target is calculated after the detection and removal of the former, i.e.,
% \begin{align}\label{eq:Pd_multi_tgt}
% P^{(2)}_{\rm d} & = \PP\left(S(\nu_{2}, \tau_{2}, \phi_{2}) > T_r(\nu_{2}, \tau_{2}, \phi_{2})  | S(\nu_{1}, \tau_{1}, \phi_{1}) > T_r(\nu_{1}, \tau_{1}, \phi_{1}) \right)
% \end{align}
% \begin{equation}
% 	P_{\rm d}=\frac{\sum_{p=0}^{P-1}P_{\rm d}(p)}{P}\,,
% \end{equation} 
%where $P_{\rm d}^{(p)}$ denotes the detection probability of the $p$-th target. 

%%%%%%%%%%%%%%%%%%%%%%%%%%%%%%%%%%%%%%%%%
\subsubsection{Tracking Mode}
\begin{figure}[h]
	\centering
	\hspace{-2cm}
	\includegraphics[scale=1.5]{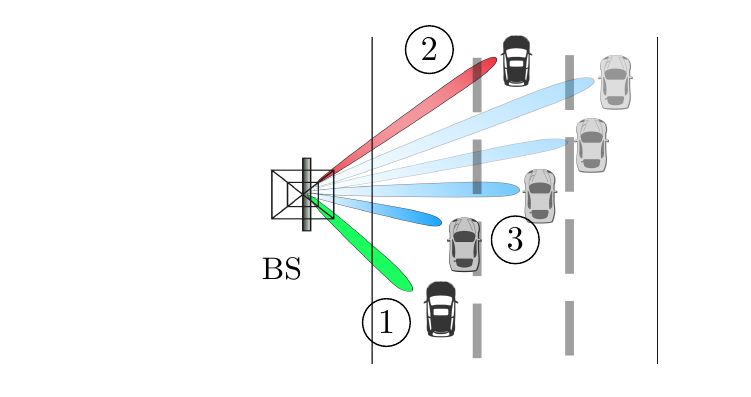}
	\caption{Simulation Scenario in Tracking mode where each of the previously acquired users receives a very narrow signal via a dedicated RF chain.}
	\label{fig:Trck_Multi_animation}
\end{figure}

\begin{figure}[h]
	\centering
	\includegraphics[scale=0.85]{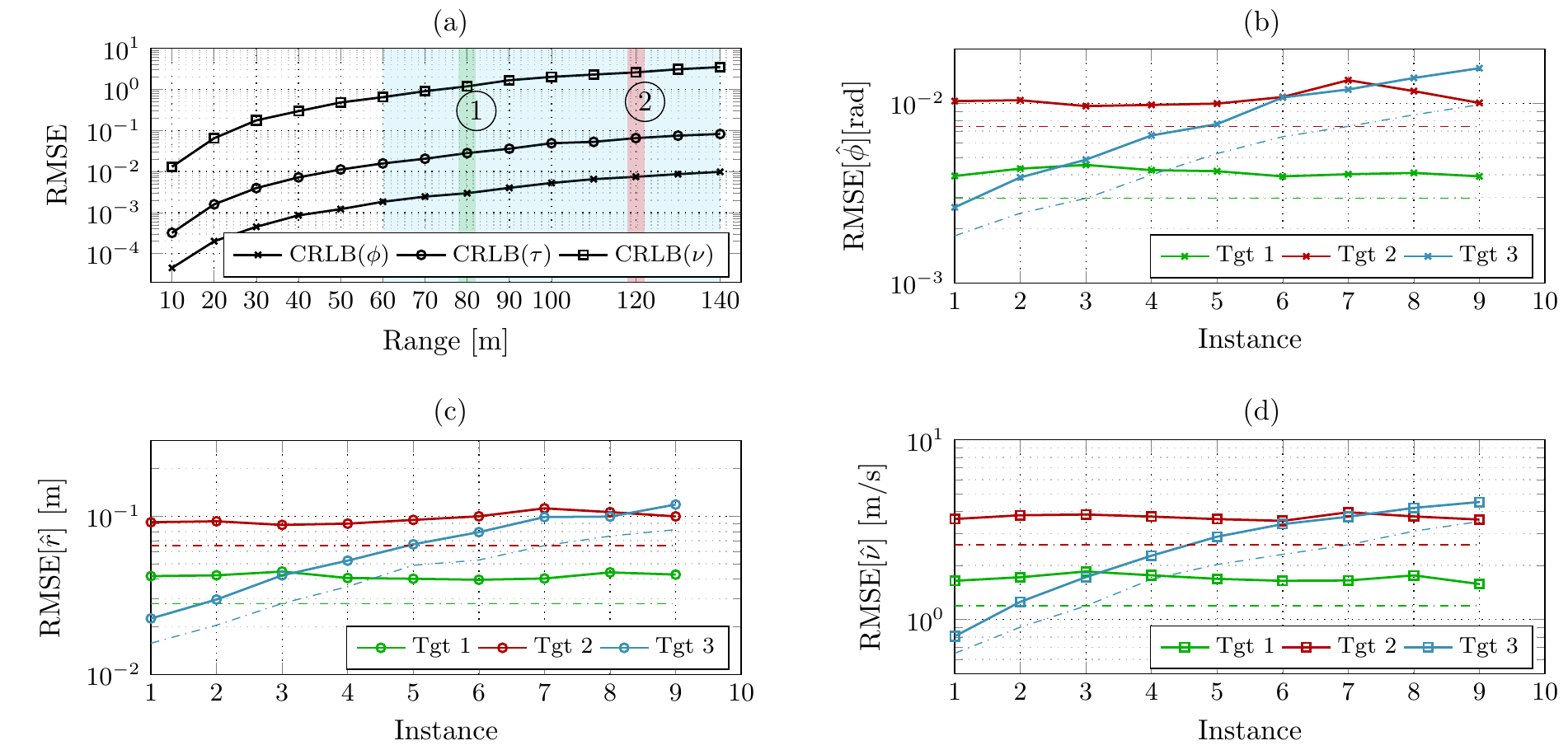}
	\caption{Estimation performance of Tracking mode. Plot (a) indicates the RMSE values of the three considered parameter estimates as dictated by the \ac{CRLB} over range (distance). The colored bars indicate the values for the two fixed targets in the setup (see Fig.~\ref{fig:Trck_Multi_animation}). Plots (b)-(d) depict estimation performance for each target, in AoA, range and radial velocity, respectively.}
	\label{fig:Est_Trck_Multi}
\end{figure}

Next, the parameter estimation performance of the Tracking mode is considered. Here, the \ac{BS} sends individually beamformed data streams to $P$ already acquired targets, i.e., {\em Users}.  This results in a significantly higher \ac{BF} gain of the transmitter and, 
consequently, a better estimation performance over a wider distance. 
As shown in Fig.~\ref{fig:Trck_Multi_animation}, in the simulated scenario three distinct users are considered where the first and second are positioned at a fixed distance and angle from the \ac{BS}, and the third user's location is changed. Notice that this does not represent relative motion: at each location of the third target, we consider fixed range, Doppler, and AoA for all targets, and perform Monte Carlo simulation of the parameter estimation scheme of Section \ref{sec:new-para-estimation}. The results are reported in Fig.~\ref{fig:Est_Trck_Multi}.

We compare the achieved \ac{RMSE} with the corresponding CRLB for all users, 
indicated by the dashed curves. 
Interestingly, when two targets have the same range, i.e., location instances 3 and 7, 
we notice a slight increase in the estimation \ac{RMSE}. This is because the targets are not distinguishable in the delay domain. 
However, thanks to their angular separation, the estimation of all the parameters (including the range) remains very accurate. 
We have also noticed by extensive simulation, that $\Am$ in \eqref{log_lkhd_1}
can be safely considered as diagonal, leading to the approximated ML estimator in 
\eqref{log_lkhd_3}. This is further confirmed by the fact that the performance of the proposed estimator follows closely the \ac{CRLB}.

% \begin{table}[h]
% \centering
% \caption{Entries of $\Am$ in eq.~\eqref{S_signal} for 3 Users}  
% % leftmost table of the top level table
% \begin{tabular}{ |c|c|c| } 
% \hline
% 1.000 &0.003 &0.007 \\
% \hline
% 0.005 &1.000  &0.004\\
% \hline
% 0.009 &0.007 &1.000 \\
% \hline
% \end{tabular} 
% \label{tab: A_3x3}% top level tables, with 2 columns
% \end{table} 

% however due to multi-dimensional computation complexity of the ML estimation, we have limited the fine resolution search as indicated by the dashed lines. 
\begin{remark} \label{remark3}
We would like to emphasize that the beam-space MIMO approach proposed in this work is not limited to a specific type of beam shape. As an example, when latency is not of concern and therefore a large value of $B$ can be considered, Fourier type beams can be used which will increase the SNR at receiver due to higher \ac{BF} gain. In other applications with relatively narrow \ac{FoV} requirements, it is possible to replace these flat-top beams or Fourier beams with a \ac{BF} codebook obtained from Slepian sequences, which provide an optimal orthogonality condition and angle concentration around the currently estimated target AOA. 
This is presented in another related work of ours \cite{BeamRefinement}. \hfill $\lozenge$
\end{remark}

\section{Conclusions}\label{sec:Conclusions}

% In this paper we propose an efficient \ac{ML}-based algorithm able to jointly perform target detection and radar parameters estimation, i.e., range, velocity, and \ac{AoA}, by using a \ac{MIMO} mono-static radar adopting on \ac{OTFS} modulation and operating in different modes. Simulation results confirm the robustness of the algorithm both in term of target identifiability, by exploiting a \ac{SIC} mechanism, and in term of estimation error. Moreover, by targeting \ac{mmWave} communications, the equipment adopts a number of \ac{RF} chains much lower w.r.t. the number of antennas. However, this limitation does not preclude to reach the theoretical performance indicated by the \ac{CRLB}, as a  further proof of the correctness of the analytical and algorithmic design. At last, even if the focus is \ac{mmWave} communication, the proposed scheme can spread the entire spectrum of frequencies, thus it is not precluded in a single operational setup. 

In this paper, we proposed a beam-space MIMO radar approach for joint data transmission and 
radar parameter estimation based on OTFS modulation and targeting \ac{mmWave} applications. 
The beam-space approach consists of reducing the $\Na$-dimensional received signal at the 
radar Rx antenna array to an $\Nrf \ll \Na$ projected observation, where the 
projection is operated by analog \ac{BF}. In this way, only $\Nrf$ RF-chains (demodulation and A/D conversion) are needed. We designed a suitable \ac{BF} codebook and proposed a multi-block detection/estimation scheme, where the projection beam patterns are changed over $B \geq 1$ blocks. 
We considered two relevant scenarios for joint communication and sensing, namely, Discovery and Tracking modes. In Discovery mode, the \ac{BS} transmits a wide angle beacon signal and aims to detect known targets (e.g., users entering the cell). In Tracking mode, the \ac{BS} transmits multiple 
individually beamformed (with narrow beams)  data streams to already acquired users, scheduled to be sufficiently separated in the angle domain. 

For Discovery mode, we proposed a sequential target detection with successive interference cancellation, able to cope with the near-far effect yielding target masking in the case of multiple targets. For Tracking mode, we proposed an approximated ML parameter estimator which has relatively low complexity and is able to approach the \ac{CRLB} on a wide range of the parameters.  
A few interesting directions are left for future work. These include the further optimization of the hybrid beamforming matrices, the comparison with other radar or/and communication waveforms, and the inclusion of such radar-aided techniques in effective schemes for initial beam acquisition (e.g., radar-enhanced initial beam alignment, for which the Discovery mode is relevant \cite{Islam}), 
and in effective schemes for beam tracking (e.g., in conjunction with mobility models and 
tracking algorithms, for which the Tracking is relevant \cite{FernandoTrack1}).

\section{Acknowledgment}
The work of Saeid K. Dehkordi has received funding from the German Federal Ministry of Education and Research within the research project ForMikro-6GKom (project number 16ES1107). The work of Lorenzo Gaudio and Giulio Colavolpe is supported by Fondazione Cariparma, under the TeachInParma Project. The work of M. Kobayashi and G. Caire is supported by the DFG, Grant agreement numbers KR 3517/11-1 and CA 1340/11-1, respectively. The authors acknowledge the financial support by the Federal Ministry of Education and Research of Germany in the program of “Souverän. Digital. Vernetzt.” Joint project 6G-RIC, project identification number: 16KISK030.

\appendices

%%%%%%%%%%%%%%%%%%%%%%%%%%%%%%%%%%%%%%%%%%%%%%%%%%%%%%%%%%%%%%%%%%%%%%%%%%%%%%%%%%%%%%%%%%
\section{Design of the beamforming vectors} \label{appendix:BF} 

Let $\fv$ be a beamforming vector of dimension $\Na$. The complex-valued (amplitude and phase) 
beam pattern radiated by the array at each sampling point $\tilde{\phi}_i,~ i\in [1,...,G]$ of a discrete angular set $\{\tilde{\Omega}\},~ (\left|\tilde{\Omega}\right| = G)$ can be calculated as the inner product of the vector $\fv$ and the array response vector $\av(\phi)$ at the given grid angle, i.e., $\av^\H(\tilde{\phi}_i)\fv$. 

The design problem of interest is to find $\fv$ to approach a desired 
radiation pattern $\bar{\bv} \in \RR^{\Na}$. The entries of $\bar{\bv} = [\bar{b}_1,...,\bar{b}_{G}]$ are magnitudes of the radiation pattern at each of the $G$ discrete angles. In particular, we fix $\bar{\bv}$ to have a constant level in a pre-determined 
angle range around the boresight direction of the array (zero angle) and such that the values
corresponding to the rejection directions (sidelobes) are below a certain threshold with respect to the maximum (center beam). 
By letting $\Am =[\av(\tilde{\phi}_1), \dots, \av(\tilde{\phi}_{G})]$, this problem can be formulated as a magnitude least-squares problem which belongs to the class of problems addressed by \cite{MLS_2, Kassakian}. 
\begin{align}
    \min_{\fv} & \quad \| \Am^\H \fv-\bar{\bv}\|^2 \nonumber\\ \label{eq: bp_optim_1}
    \text{s.t.} & \quad  \fv^\H \Am^\H \Am \fv = 1
\end{align}
%\begin{equation} \label{eq: bp_optim_1}
%\begin{split}
%\mathcal{P}(1):\quad &\underset{\fv}{\min} %\quad\left\{\sum^{N_a}_{j=1}(\vert \matr{A}^{T}_{:,j}\fv\vert -\bar{\bv}_{j})^{2}\right\} \\
%& \text{s.t.}\quad\quad \fv^{H}\Am^{H}\Am\fv=1 %p
%\end{split}
%\end {equation}  
where the constraint in \eqref{eq: bp_optim_1} imposes unit transmit power. 
Problem \eqref{eq: bp_optim_1} can be solved as a semidefinite relaxation of the magnitude least-squares problem \cite{Kassakian}.
Depending on the operating scenario, a beam pattern can focus the transmitted energy on a certain given angular sector (i.e., \ac{FoV} equal to $\Omega$). In order to define our design in a flexible manner, the \ac{FoV} is divided into a central section $\Omega_m$ covering $G_m$ discrete directions each with magnitude of $\sigma_m$ and the remaining sections (modulo the interval $[-\pi,\pi]$ denoted by $\Omega_p$, with $G_p$ grid points and magnitude $\sigma_p)$.  The desired beam pattern $\bar{\bv}$ has a total power of $G_m\sigma_m^2$  in the central sector and $G_p\sigma_p^2 = 1-G_m\sigma_m^2$ in the remaining peripheral sections. By controlling the main and peripheral sections, we can control the width of the main lobe and 
the side lobes rejection. 

\begin{figure}
	\centering
	\includegraphics[scale=0.8]{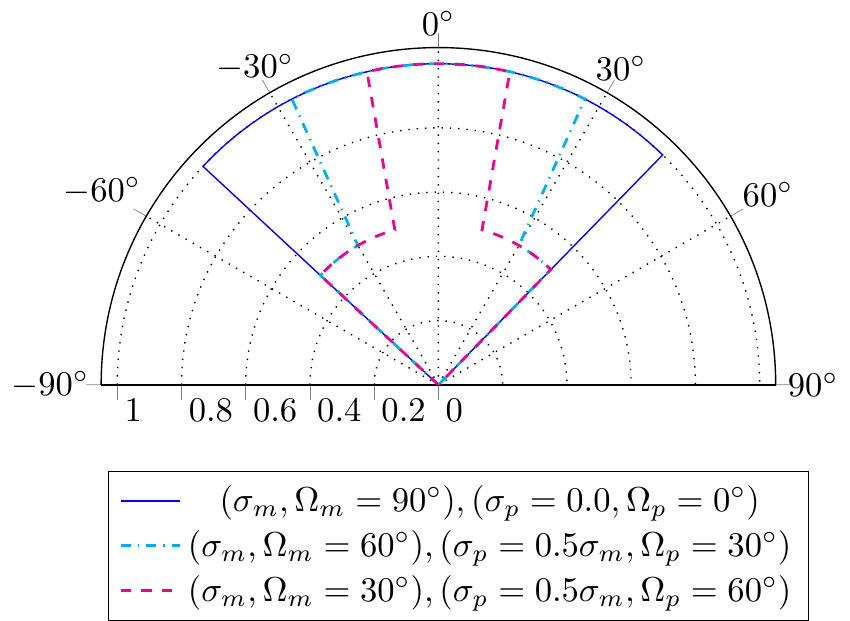}
	\caption{Examples of beampattern masks with varying main lobe widths and sidelobe levels.}
	\label{fig:TX_Mask-Na64}
\end{figure}

\begin{figure}
	\centering
	\includegraphics[scale=0.75]{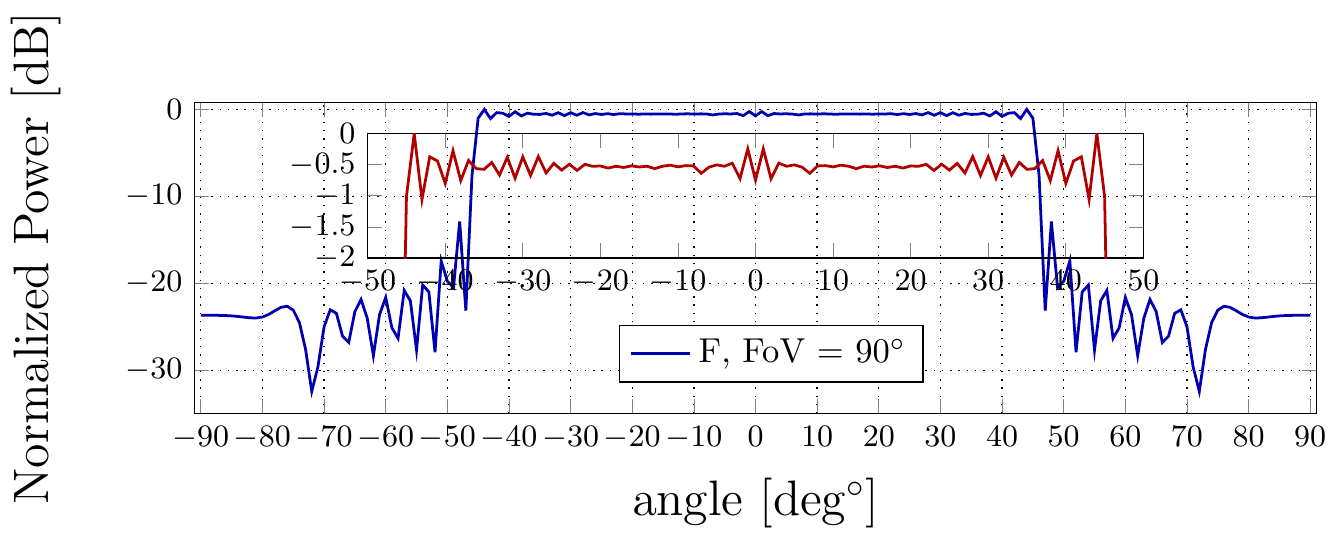}
	\caption{Tx beampattern for a \ac{FoV} of $90^{\circ}$ and $\Na=64$. 
	The zoomed-in plot depicts the minimal ripple within the main beam.}
	\label{fig:TX_BF90_Na64_zoom}
\end{figure}

Fig.~\ref{fig:TX_Mask-Na64} shows a few examples of the used design beamforming masks. 
Fig.~\ref{fig:TX_BF90_Na64_zoom} illustrates the achieved flat-top beampattern for an \ac{FoV} of $90^{\circ}$, corresponding to the wide Tx beam used in Discovery mode in this paper.

%%%%%%%%%%%%%%%%%%%%%%%%%%%%%%%%%%%%%%%%%%%%%%%%%%%%%%%
\section{Adaptive Threshold for Detection }\label{CFAR}

\newcommand{\ind}[1]{\mathds{1}_{\left\lbrace #1 \right\rbrace}}
\newcommand{\norm}[1]{\left|\left| #1 \right|\right|}
\newcommand{\bigO}[1]{\ensuremath{\mathop{}\mathopen{}O\mathopen{}\left(#1\right)}}
\newcommand{\smallO}[1]{\ensuremath{\mathop{}\mathopen{}o\mathopen{}\left(#1\right)}}

With reference to the notation introduced in Section \ref{disc_paramEst}, 
the decision statistics at each Doppler-delay-angle bin $(\nu, \tau, \phi)\in \Gamma \times \widehat{\Omega}$ under hypothesis $\Hc_0$ if given by 
\begin{align}
S^{\Hc_0}(\nu, \tau, \phi) &:= \frac{\left | \sum_{b=1}^B \underline{\wv}_b^\herm \underline{\Gm}_b(\nu, \tau, \phi) \underline{\xv}_b \right |^2}{\sum_{b=1}^B \| \underline{\Gm}_b (\nu, \tau, \phi)\underline{\xv}_b \|^2},~~
S^{\Hc_1}(\nu, \tau, \phi):= \frac{\left | \sum_{b=1}^B \underline{\yv}_b^\herm \underline{\Gm}_b(\nu, \tau, \phi) \underline{\xv}_b \right |^2}{\sum_{b=1}^B \| \underline{\Gm}_b (\nu, \tau, \phi)\underline{\xv}_b \|^2}.\
\label{S-under-H0}
\end{align}
We wish to compute an adaptive thresholds $T_r(\nu, \tau, \phi)$ to compare 
the decision statistic $S(\nu,\tau,\phi)$ and decide for $\Hc_0$ or $\Hc_1$, 
such that a target false alarm probability is achieved. 

Notice that the function $S^{\Hc_0}(\nu, \tau, \phi) $ is the squared magnitude of a Gaussian complex circularly symmetric random obtained as the linear projection of the \ac{AWGN} vector. Hence, it is exponentially distributed (for given $\{\underline{\Gm}_b(\nu, \tau, \phi) \underline{\xv}_b\}_{b=1}^B$).

 \begin{figure}
	\centering
	\hspace{3cm}
	\includegraphics[scale=1.0]{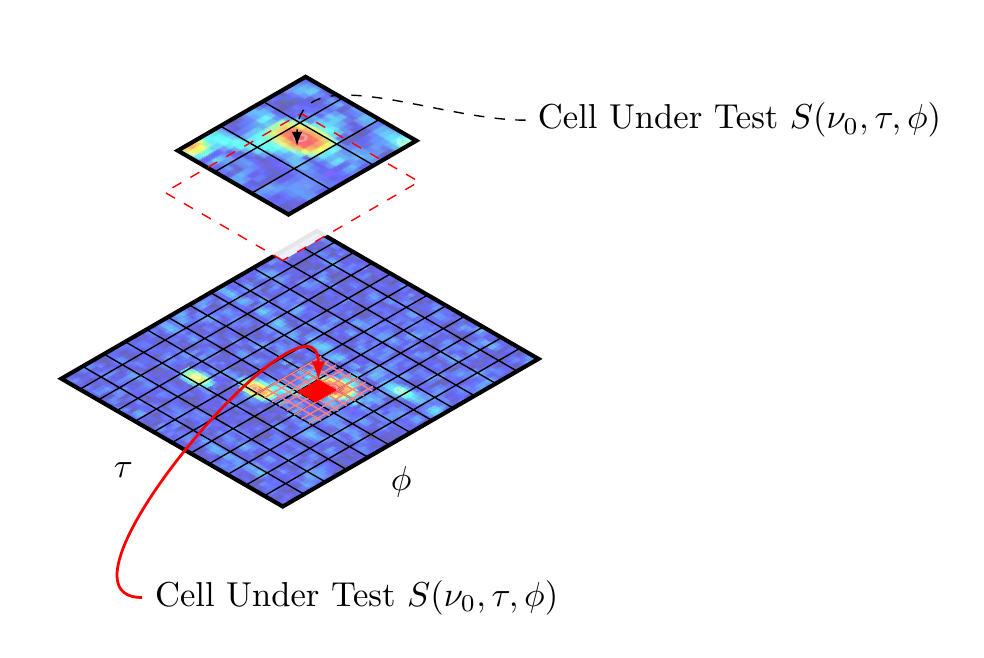}
	\caption{Graphical representation of \ac{OS-CFAR} windowing. For a plane cut at $\nu=\nu_0$, the window is shifted around each location $(\nu_0,\tau,\phi)$ in the search space.}
	\label{fig:Cfar_schematic}
\end{figure}

We can estimate the distribution locally at each Doppler-delay-angle bin $(\nu, \tau, \phi)\in \Gamma \times \widehat{\Omega}$. 
Following the OS-\ac{CFAR} procedure \cite[Chapter 6.5]{richards2014fundamentals}, 
we first define a set of neighboring bins, denoted by $\Cc(\nu, \tau, \phi)$, centered at $(\nu,\tau,\phi)$ (see Fig.~\ref{fig:Cfar_schematic}). Let $N_c = \max_{(\nu, \tau, \phi)\in \Gamma \times \widehat{\Omega}|\Cc(\nu, \tau, \phi)|}$ denote the size of the neighboring bins 
for most of bins in the search space (notice that for some bins in the boundary of the domain the size of the neighboring set $\Cc(\nu, \tau, \phi)$ may be less than $N_c$). 
For each Doppler-delay-angle bin $(\nu, \tau, \phi)\in \Gamma \times \widehat{\Omega}$, 
we evaluate the value of $S(\nu', \tau', \phi')$ for $(\nu', \tau', \phi')\in \Cc (\nu, \tau, \phi)$ and sort them in an increasing order such that
% \begin{align}
% S^{\Hc_0}(\nu_1, \tau_1, \phi_1) \leq S^{\Hc_0}(\nu_2, \tau_2, \phi_2) \leq  \dots \leq S^{\Hc_0}(\nu_{N_c}, \tau_{N_c}, \phi_{N_c})
% \end{align}
\begin{align}
S(\nu_1, \tau_1, \phi_1) \leq S(\nu_2, \tau_2, \phi_2) \leq  \dots \leq S(\nu_{N_c}, \tau_{N_c}, \phi_{N_c})
\end{align}
where $(\nu_i, \tau_i, \phi_i) \in \Cc(\nu, \tau, \phi)$ denotes the $i$-th element in the above ordered statistics of the neighboring set. 
Assuming that no target falls in the neighboring set 
$\Cc(\nu, \tau, \phi)$, the above ordered statistics yields an empirical 
cumulative distribution function (CDF) of $S(\nu, \tau, \phi)$. Hence, 
the threshold can be determined by choosing a given percentile $\kappa$ of this empirical CDF, 
and scaling it by a factor $\alpha$ that depends on the specific problem at hand and must be tuned simulation. Specifically, fixing $\kappa\in (0, 1)$,  
we express the adaptive threshold $T_r(\nu, \tau, \phi)$ as
\begin{align}
T_r(\nu, \tau, \phi) = \alpha S(  \nu_{\lceil \kappa N_c \rceil}, 
\tau_{\lceil \kappa N_c \rceil}, \phi_{\lceil \kappa N_c \rceil}).
\end{align}
The corresponding false alarm probability at bin $(\nu, \tau, \phi)$ if given by 
\begin{align}\label{eq:Pfa}
P_{\rm fa}(\nu, \tau, \phi)  & = \PP(S(\nu, \tau, \phi) > T_r(\nu, \tau, \phi)  | \Hc_0) \nonumber \\
&\approx 1 - \hat{F}_{S^{\Hc_0}}(T_r(\nu, \tau, \phi))
\end{align}
where $\hat{F}_{S^{\Hc_0}}$ denotes the empirical CDF 
of $S^{\Hc_0}(\nu, \tau, \phi)$ calculated from the ordered statistics defined above. 
The average false alarm probability is given by 
\begin{align}
\bar{P}_{\rm fa} = \frac{1}{|\Gamma| |\widehat{\Omega}| } \sum_{(\nu, \tau, \phi)\in \Gamma \times \widehat{\Omega} }P_{\rm fa}(\nu, \tau, \phi) 
\end{align}
and depends on $\kappa$ and $\alpha$. Then,  we can tune these parameters such that a target average false alarm probability is satisfied. This is obtained by simulation (in our case) or by training (in a real-world scenario). 

\begin{figure}
	\begin{subfigure}{.25\columnwidth}
		\centering
		\includegraphics[scale=0.2]{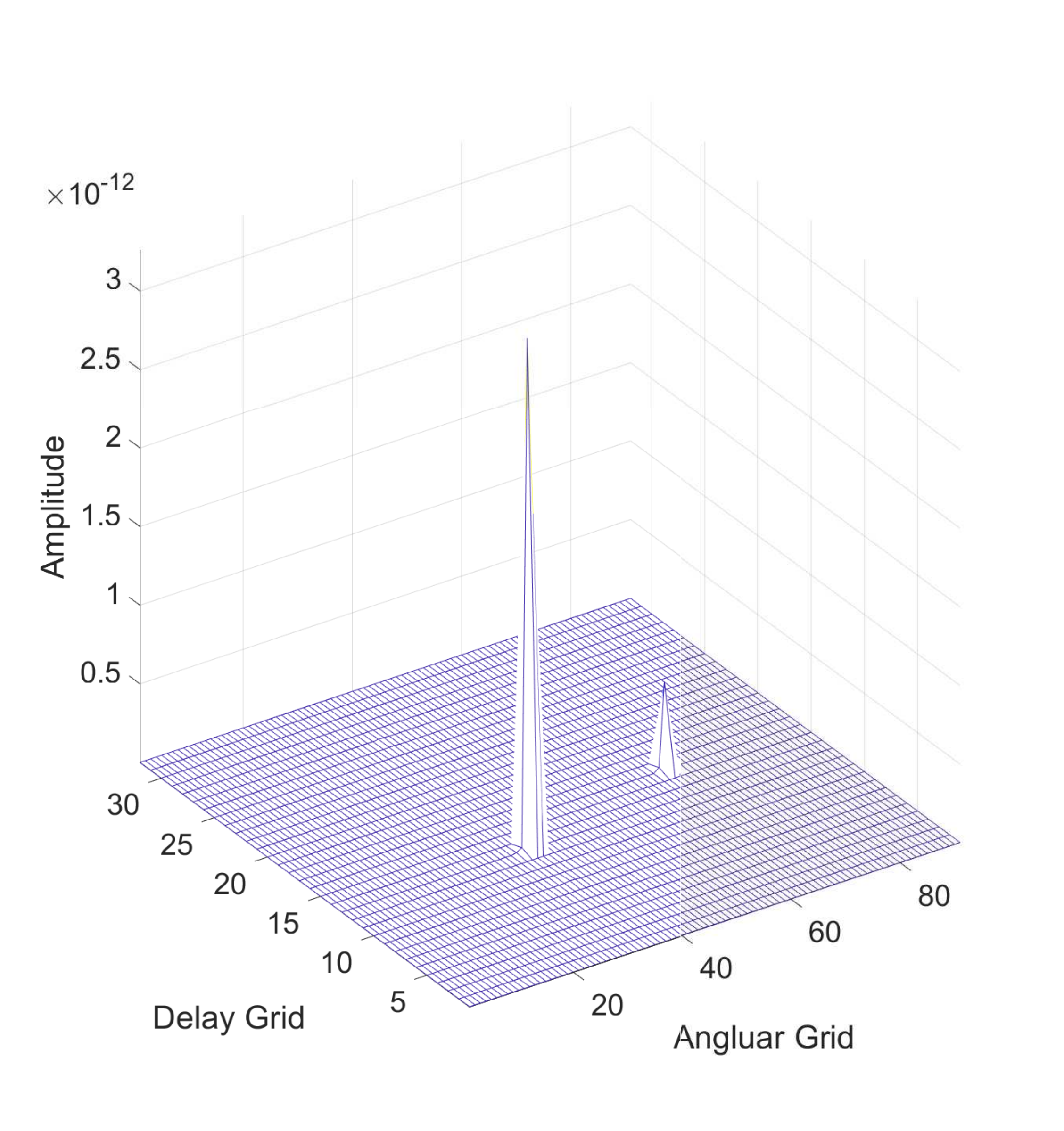}
		\caption{Signal}
		\label{fig: cfar_procedure_a}
	\end{subfigure}
	\begin{subfigure}{.25\columnwidth}
		\centering
		\includegraphics[scale=0.25]{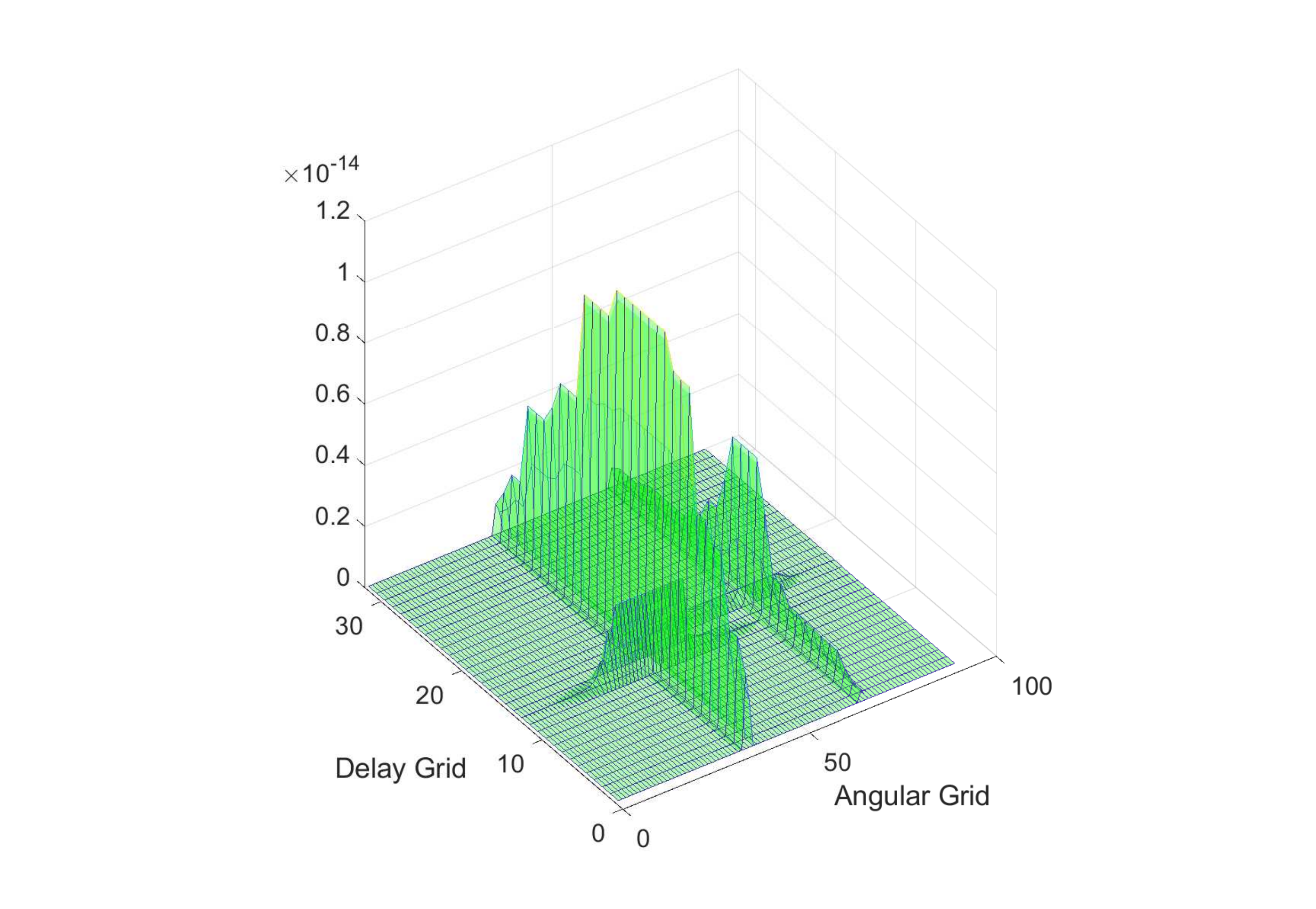}
		\caption{Threshold}
		\label{fig: cfar_procedure_b}
	\end{subfigure}
	\begin{subfigure}{.28\columnwidth}
	%	\centering
	\vspace{1cm}
	    \hspace*{2cm}
		\includegraphics[scale=0.2]{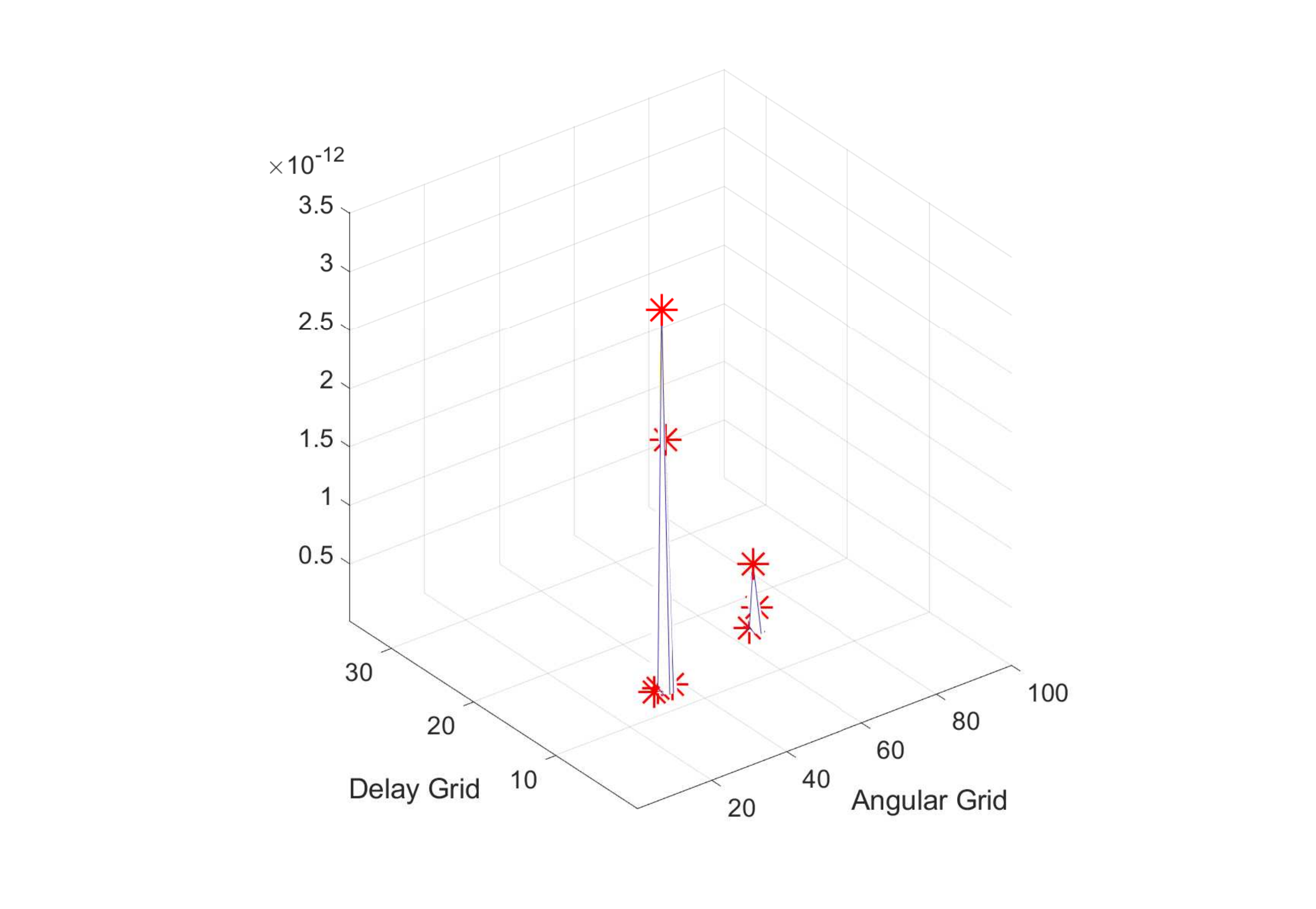}
		\caption{Threshold-passed Signal points}
		\label{fig: cfar_procedure_c}
	\end{subfigure}
	\caption{Illustrative description of the three stages of the \ac{OS-CFAR} procedure, namely, (a) Received signal as input, (b) calculating the threshold and, (c) applying the threshld against the signal to obtain detections.}
	\label{fig: cfar_procedure}
\end{figure}

Fig.~\ref{fig: cfar_procedure} shows the three stages of the \ac{OS-CFAR} procedure in a 2-dimensional \textit{delay-angle} space (a depiction in 2-D space is shown for visualization clarity). Fig.~\ref{fig: cfar_procedure} (a) shows the decision statistics $S(\nu,\tau,\phi)$ for $\nu = 0$, in the delay-angle plane.
Fig.~\ref{fig: cfar_procedure} (b) shows the threshold $T_r(\nu, \tau, \phi)$ calculated according to the above OS-CFAR procedure. Fig.~\ref{fig: cfar_procedure} (c) shows the 
portion of the detection statistics in (a) above the adaptive threshold in (b). 
This determines the set of points in the grid above threshold $\Tc$ as defined in \eqref{set-T}. 
Finally, the target is identified by taking the position of the maximum of $S(\nu,\tau,\phi)$ for 
over the set $\Tc$, if this is not empty, otherwise hypothesis $\Hc_0$ (no target) is declared.

%\begin{figure}
%		\centering
%		\includegraphics[scale=0.25]{Figures/CFAR_COMPLETE.png}
%		\caption{Signal}
%		\label{fig: cfar_procedure_a}
%	\caption{CFAR Stages}
%	\label{fig: cfar_procedure}
%\end{figure}

%%%%%%%%%%%%%%%%%%%%%%%%%%%%%%%%%%%%%%%%%%%%%%%%%%%%%%%%%%%%%%%%%%%%%%%%%%%%%%%%%%%%%%%%%%%%%%%%%%%

\section{Derivation of the multi-block \ac{CRLB}}\label{app:CRLB} 
%For the sake of notation simplicity, we replace $\mathring{\thetav}$ with %$\thetav$ in the Appendix.  
In order to calculate the multi-block \ac{CRLB}, we replace $\Psi_{n, k,m,l}(\nu,\tau)$ defined in \eqref{eq:Psi-Mat} with its 
approximated version $\bar{\Psi}_{n, k, m, l}(\nu,\tau)$, i.e. the cross-talk between Doppler-delay bins $(n, m)$ and $(k, l)$ where $n, k\in [0:N-1]$ and $m, l\in [0:M-1]$ given by 
\begin{align}%\label{eq:Psi-Mat}
%	\Psi_{n, k,m,l}(\tau_p, \nu_p) & =\sum_{k', n', l', m'} \frac{C_{g_{\rm rx}, g_{\rm tx}}((k'-n')T -\tau_p, (l'-m') \Delta %f-\nu_p)}{NM}  e^{j 2\pi n' T \nu_p} e^{-j2\pi l' \Delta f  \tau_p}e^{j 2\pi \left(\frac{n'k}{N}- \frac{m' l}{M}\right)}e^{-j %2\pi\left(\frac{k' n}{N}-\frac{l' m}{M}\right)}  \nonumber \\
%	&\approx 
\bar{\Psi}_{n, k, m, l} (\nu, \tau)  \eqdef	 \frac{1}{NM} \sum_{n'=0}^{N-1}\alpha_{n, k,n'} (\nu) \sum_{m'=0}^{M-1} \beta_{m',k, m, l} ( \nu, \tau) 
\end{align} 
where we defined
\begin{subequations}
\begin{align}
\alpha_{n, k, n'} (\nu) &= e^{j2\pi  (k-n+\nu  NT)\frac{n'}{N}} \\
\beta_{m',k,m,l} (\tau, \nu) &= e^{j2\pi  (m-m'+\tau M\Delta f) \frac{m'}{M}} e^{j2\pi \nu \frac{l}{M\Delta f}}
	\begin{cases}
	\begin{array}{ll}
	1 & l\in \Lc_{\mathrm{ICI}}(\tau):= [0, M-l_{\tau}-1]\\
	e^{-j2\pi\left(\nu T+\frac{k}{N}\right)} & l\in \Lc_{\mathrm{ISI}}(\tau) :=[M-l_{\tau},M-1].\\
	\end{array}
	\end{cases}
\end{align}
\end{subequations}
where we let $l_{\tau}= \lceil \frac{\tau}{T/M}\rceil$.  For the later use, we also calculate the derivative w.r.t. $(\tau, \nu)$. 
\begin{subequations}
\begin{align}
\frac{\partial \alpha_{n, k, n'} (\nu)}{\partial \nu} &= (j2\pi n' T)  \alpha_{n, k, n'} (\nu) \label{eq:alpha_nu}\\ %e^{j2\pi  (k-n+\nu  NT)\frac{n'}{N}} \\
\frac{\partial\beta_{m',k,m,l} ( \nu,\tau)}{\partial \nu} &=  \beta_{m',k,m,l} (\tau, \nu)%e^{j2\pi  (m-m'+\tau M\Delta f) \frac{m'}{M}} e^{j2\pi \nu \frac{l}{M\Delta f}}
	\begin{cases}
	\begin{array}{ll}
j2\pi  \frac{l}{M\Delta f} & l\in \Lc_{\mathrm{ICI}}(\tau)\\
	j2\pi  \left(\frac{l}{M\Delta f} -T \right) & l\in \Lc_{\mathrm{ISI}} (\tau)\\
	\end{array}
	\end{cases} \nonumber\\
	& = j 2 \pi g(l)  \beta_{m',k,m,l} (\nu,\tau) \label{eq:beta_nu}\\
\frac{\partial\beta_{m',k,m,l} (\nu,\tau)}{\partial \tau} &=  (j2\pi m' \Delta f) \beta_{m',k,m,l} ( \nu,\tau) \label{eq:beta_tau}
\end{align}
\end{subequations}
where we let $g(l) = \frac{l}{M\Delta f} $ for $l\in \Lc_{\rm ICI}(\tau)$ and $g(l) = \frac{l}{M\Delta f} -T$ for $l\in \Lc_{\rm ISI}(\tau)$.

For the notation simplicity, we also define
\begin{align}
\alphav_{n, k} (\nu) & =[\alpha_{n, k,0} (\nu), \dots, \alpha_{n, k,N-1} (\nu) ]^\T  \\
\betav_{k,m,l} (\nu,\tau) &= [\beta_{0,k,m,l} ( \nu,\tau) , \dots, \beta_{M-1,k,m,l} ( \nu,\tau) ]^\T 
\end{align}
Hereafter, we use the following approximated expression of the desired signal vector. 
\begin{align}
\bar{\sv}_{b, n, m}(\thetav) &= A e^{j \psi} \Um_{b}^\H \av(\phi)\av^\H (\phi)\fv \sum_{k=0}^{N-1}\sum_{l=0}^{M-1}\bar{\Psi}_{n, k, m, l}( \nu,\tau) x_b[k, l] \nonumber\\
&=\frac{A}{NM} e^{j \psi} \Um_{b}^\H \av(\phi)\av^\H (\phi)\fv \sum_{k=0}^{N-1}\sum_{l=0}^{M-1}\onev_N^\T  \alphav_{n, k} (\nu) \onev_M^\T \betav_{k,m,l} (\nu,\tau)  x_b[k, l]
\end{align}

\noindent {\bf Derivatives w.r.t. delay and Doppler shift~~}  \\
We remark that $( \nu, \tau)$ appear only in $\bar{\Psi}_{n, k,m,l}( \nu,\tau)$. 
From \eqref{eq:beta_tau}, we have
\begin{align}
	\frac{\partial\bar{\Psi}_{n, k, m, l}}{\partial\tau}
%	&= \frac{1}{NM}\sum_{n'} e^{j2\pi\left(k-n+\nu_p NT\right)\frac{n'}{N}}\sum_{m'} e^{j2\pi  (l-m+\tau_p M\Delta f) \frac{m'}{M}} (j2\pi m'\Delta f)e^{j2\pi\nu_p\frac{l}{M\Delta f}}\times
%	\begin{cases}
%	\begin{array}{ll}
%	1 & l_{\mathrm{ICI}}\\
%	e^{-j2\pi\left(\nu_p T+\frac{k'}{N}\right)} & l_{\mathrm{ISI}} \\
%	\end{array}
%	\end{cases}\, \\
	%&=  \frac{j2\pi \Delta f}{NM} \sum_{n'} \alpha_{n, k,n'}(\nu_p) \sum_{m'} m' \beta_{k, m, l,m'} (\tau_p, \nu_p)\nonumber\\
	&=\frac{j2\pi \Delta f}{NM} \onev_N^\T \alphav_{n, k}(\nu)  \cv_M^\T\betav_{k, m, l} (\nu, \tau)
\end{align}
where we let $\cv_M^\T= [0, 1, \dots, M-1]$. 

From \eqref{eq:alpha_nu} and \eqref{eq:beta_nu}, we have 
\begin{align}
\frac{\partial \bar{\Psi}_{n,k, m, l}}{\partial \nu} &=  \frac{1}{NM} 
 \Bigg[ \sum_{n'=0}^{N-1}\frac{\partial \alpha_{n, k,n'} (\nu)}{\partial \nu} \sum_{m'=0}^{M-1} \beta_{m',k, m, l} (\nu, \tau)
 + \sum_{n'=0}^{N-1}\alpha_{n, k,n'} (\nu) \sum_{m'=0}^{M-1} \frac{\partial \beta_{m',k, m, l} (\nu, \tau)}{\partial \nu} \Bigg] \nonumber \\
% &= \frac{j2\pi}{NM} 
% \Bigg[ T\sum_{n'=0}^{N-1} n'  \alpha_{n, k, n'} (\nu_p) 
%  \sum_{m'=0}^{M-1} \beta_{m',k, m, l} (\tau_p, \nu_p)
% + \sum_{n'=0}^{N-1}\alpha_{n, k,n'} (\nu_p) \sum_{m'=0}^{M-1}  g(l)  \beta_{m',k,m,l} (\tau, \nu)
% \Bigg]  \nonumber\\
 &= \frac{j2\pi}{NM} 
 \Bigg[ T \cv_N^\T  \alphav_{n, k} (\nu) \onev_M^\T \betav_{k, m, l} ( \nu,\tau)
 +  g(l)   \onev_N^\T \alphav_{n, k} (\nu) \onev_M^\T \betav_{k,m,l} (\nu, \tau)
 \Bigg] 
\end{align}
These yield the following derivatives
\begin{align}\label{eq:tau}
\frac{\partial \bar{\sv}_{b,n,m}}{\partial \tau} &=  \frac{j2\pi \Delta f}{NM}  A e^{j \psi} \Um_{b}^\H \av(\phi)\av^\H (\phi)\fv 
\sum_{k, l} \onev_N^\T \alphav_{n, k}(\nu)  \cv_M^\T\betav_{k, m, l} (\nu, \tau)  x_b[k, l] \\ \label{eq:nu}
\frac{\partial \bar{\sv}_{b,n,m}}{\partial \nu} %&= \onev\{q=p\} \frac{j2\pi}{NM}  A_p e^{j \psi_p} \Um_{b}^\H \av(\phi_p)\av^\H (\phi_p)\fv  \sum_{k, l}\Bigg[ T\sum_{n'=0}^{N-1} n'  \alpha_{n, k, n'} (\nu_p) 
%  \sum_{m'=0}^{M-1} \beta_{m',k, m, l} (\tau_p, \nu_p) \nonumber\\
%&~~~~~~~~~~~~ + \sum_{n'=0}^{N-1}\alpha_{n, k,n'} (\nu_p) \sum_{m'=0}^{M-1}  g(l)  \beta_{m',k,m,l} (\tau, \nu)
% \Bigg] x_{k, l}\\
 &= \frac{j2\pi}{NM}  A e^{j \psi} \Um_{b}^\H \av(\phi)\av^\H (\phi)\fv  \sum_{k, l}  x_b[k, l] d_{k, l} 
 %\nonumber\\
 %&\;\;\;\;\;\;\;\;\; \Bigg[ T \cv_N^\T  \alphav_{n, k} (\nu_p) \onev_M^\T \betav_{k, m, l} (\tau_p, \nu_p)
 %+  g(l)   \onev_N^\T \alphav_{n, k} (\nu_p) \onev_M^\T \betav_{k,m,l} (\tau_p, \nu_p)
 %\Bigg] 
\end{align}
where 
\begin{align}\label{eq:d}
d_{k, l}(\nu, \tau)=  T \cv_N^\T  \alphav_{n, k} (\nu) \onev_M^\T \betav_{k, m, l} (\nu,\tau)
 +  g(l)   \onev_N^\T \alphav_{n, k} (\nu) \onev_M^\T \betav_{k,m,l} (\nu, \tau)
\end{align}

\noindent {\bf Derivatives w.r.t. complex channel coefficients}  
\begin{align}\label{eq:coefficient}
\frac{\partial \bar{\sv}_{b,n, m}}{\partial A } &= e^{j \psi} \Um_{b}^\H \av(\phi)\av^\H (\phi)\fv ~%\xsf_{b,n,m}(\tau_p, \nu_p) 
\sum_{k=0}^{L-1} \sum_{l=0}^{M-1} \bar{\Psi}_{n, k}[ m, l] x_b[k, l]  \\
\frac{\partial \bar{\sv}_{b, n, m}}{\partial \psi } &= j A e^{j \psi} \Um_{b}^\H \av(\phi)\av^\H (\phi)\fv %~\xsf_{b,n,m}(\tau_p, \nu_p) 
\sum_{k=0}^{L-1} \sum_{l=0}^{M-1} \bar{\Psi}_{n, k}[ m, l] x_b[k, l] 
\end{align}

\noindent{\bf Derivatives w.r.t. AoA~~}
  \begin{align}\label{eq:AoA}
\frac{\partial \bar{\sv}_{b,n,m}}{\partial \phi }&=  j \pi \cos(\phi) A e^{j \psi}\Um_{b}^\H  \av(\phi)\av^\H(\phi) \odot \Bm  \fv %~\xsf_{b,n,m}(\tau_p, \nu_p) 
\sum_{k=0}^{L-1} \sum_{l=0}^{M-1}\bar{\Psi}_{n, k}[ m, l] x_b[k, l]
\end{align}
where $\odot$ denotes the element-wise multiplication and 
$\Bm$ is a $\Na \times \Na$ matrix whose $(m, m')$-th entry is given by 
\begin{align}
[\Bm]_{m, m'}= (m-m') , \forall m, m'\in [0:\Na-1].
\end{align}

Plugging \eqref{eq:tau}, \eqref{eq:nu}, \eqref{eq:coefficient}, and \eqref{eq:AoA} into \eqref{eq:Fisher}, we can construct a $5\times 5$ matrix:
\begin{align}
\Id(\thetav) = \left[\begin{matrix}
I_{AA} & I_{A \psi} & I_{A \phi} & I_{A \tau} & I_{A \nu} \\
I_{\psi A} & I_{\psi  \psi} & I_{\psi  \phi}  &  I_{\psi  \tau} & I_{\psi  \nu} \\
I_{\phi A} & I_{\phi  \psi} & I_{\phi  \phi}  &  I_{\phi  \tau} & I_{\phi  \nu} \\
I_{\tau A} & I_{\tau  \psi} & I_{\tau  \phi} & I_{\tau  \tau} & I_{\tau  \nu} \\
I_{\nu A} & I_{\nu  \psi} & I_{\nu  \phi}  & I_{\nu  \tau} & I_{\nu  \nu} \\
\end{matrix}\right]
\end{align}
Notice that each block is of dimension $P\times P$ and diagonal due to the indicator function appearing in \eqref{eq:tau}, \eqref{eq:nu}, \eqref{eq:coefficient}, and \eqref{eq:AoA}.

In what follows, we provide explicitly the $5\times 5$ Fisher information matrix by repeatedly 
 using the following properties.
\begin{itemize}
\item[1)]  Quadratic form with trace:
\begin{align}\label{eq:trace}
( \Um_{b}^\H \av(\phi)\av^\H (\phi)\fv)^\H( \Um_{b}^\H \av(\phi)\av^\H(\phi) \odot \Bm\fv) 
%&= \trace( \fv  \fv^\H \av(\phi_p)\av^\H (\phi_p) \Um_b \Um_b^\H \av(\phi_p)\av^\H(\phi_p) \odot \Bm) \nonumber\\
&= \left\lVert \Um_b^\H \av(\phi) \right\rVert_2^2 \fv^\H \av(\phi)  \av^\H(\phi) \odot \Bm \fv 
%&= \left\lVert \Um_b^\H \av(\phi_p) \right\rVert_2^2\  \trace \left( \fv \fv^\H \av(\phi_p) \av^\H(\phi_p) \odot \Bm \right)  
\end{align}
%where the last equality follows by recalling that the diagonal elements of $\Bm$ are all zero.
\item[2)] i.i.d. assumption on the symbols: for any block $b$, 
\begin{align}\label{eq:iidsymbol}
\EE[x_b[k, l]^*x_b[k', l']] =0, \forall (k', l') \neq (k, l), \;\;\; \EE[|x_b[k, l]|^2]=\Pav.
\end{align}
\end{itemize}
We have
\begin{align}
I_{AA} %&= \frac{2\Na}{N_0} \sum_{n, m} \left|\sum_{k, l} \Psi^0_{n,k}[m,l,t] x_{k,l}\right|^2  \nonumber\\
&= \frac{2}{\sigma_w^2} \sum_{b=1}^B  \| \Um_{b}^\T \av(\phi)\av^\H (\phi)\fv\|^2 \sum_{n=0}^{N-1}\sum_{m=0}^{M-1} 
\EE \left| \sum_{k=0}^{L-1} \sum_{l=0}^{M-1} \bar{\Psi}_{n, k,m,l} x_b[k, l] \right|^2  \nonumber\\
&\stackrel{(a)}\approx \frac{2\Pav}{\sigma_w^2} \sum_{b=1}^B \| \Um_{b}^\T \av(\phi) \av^\H (\phi)\fv\|^2 \sum_{n=0}^{N-1}\sum_{m=0}^{M-1} 
\sum_{k=0}^{L-1} \sum_{l=0}^{M-1}  |\bar{\Psi}_{n, k,m,l} |^2 
%I^{0}_{A\theta} &= \frac{2}{N_0}\Re\left\{\sum_{n, m,t} j \left|\sum_{k, l} \Psi^0_{n,k}[m,l,t] x_{k,l}\right|^2\right\} =0 \\
%I^{0}_{A \tau}&=\frac{2}{N_0} \Re\left\{ \sum_{n, m,t} e^{-j\theta_0} a_0e^{j\theta_0} \sum_{k,l} ({\Psi^0}_{n, k}[m, l,t])^\herm x^*_{k,l}  \sum_{k,l} \frac{\partial\Psi^0_{n, k}[ m, l]}{\partial\tau_0}x_{k, l} \right\} \\
%&= \frac{2}{N_0} \Re\left\{ \sum_{n, m} a_0 \sum_{k,l} 
% \frac{x^*_{k,l} }{NM} \sum_{n'} {\alpha^p}^*_{n, k,n'} (\nu_p) \sum_{m'} {\beta^p}^*_{k, m, l,m'} (\tau_p, \nu_p)
% \sum_{k,l} 
%\frac{j2\pi x_{k,l}}{NM}\sum_{m'} m' \beta^p_{k, m, l} (\tau_p, \nu_p) \sum_{n'} (n'T + g(l) ) \alpha^p_{n, k,n'}(\nu_p)
%\right\}  \\
\end{align}
where (a) follows from the i.i.d. symbols and the independence between $\{\bar{\Psi}_{n, k,m,l}\} $ and $\{x_b[k,l]\}$. 
%We remark that the term $I_{AA} $ is multiplied by $\sum_{b, t}  \left| \uv_{b, t}^\T \av(\phi_p)\av^\H (\phi_p)\fv\right|^2$ compared to the single-antenna case  i.e. the SNR increases with the number of RF chains $\Nrf$ and the number $B$ of blocks. 
Using the same argument, other diagonal elements are given by
 \begin{align} 
  I_{\psi  \psi} 
  %&=   \frac{2 A^2_0}{\sigma_w^2}  \sum_{b=1}^B \| \Um_{b}^\T \av(\phi_p)\av^\H (\phi_p)\fv\|^2 \sum_{n=0}^{N-1}\sum_{m=0}^{M-1}  \EE\left[
% \left| \sum_{k=0}^{L-1} \sum_{l=0}^{M-1} \bar{\Psi}_{n, k,m,l} x_{k, l}  \right|^2\right]\nonumber \\
 &\approx \frac{2 A^2 \Pav}{\sigma_w^2}  \sum_{b=1}^B \| \Um_{b}^\T \av(\phi)\av^\H (\phi)\fv\|^2 \sum_{n=0}^{N-1}\sum_{m=0}^{M-1}  
 \sum_{k=0}^{L-1} \sum_{l=0}^{M-1}  \left|\bar{\Psi}_{n, k,m,l} \right|^2 \\
 I_{\phi  \phi}
 %&= \frac{2  \pi^2 \cos(\phi)^2 A_0^2}{\sigma_w^2} \sum_{b=1}^B  \| \Um_{b}^\H  \av(\phi)\av^\H (\phi)\odot \Bm \fv \|^2 %\sum_{n=0}^{N-1}\sum_{m=0}^{M-1}   
 %\EE \left[\left| \sum_{k=0}^{L-1} \sum_{l=0}^{M-1}\bar{\Psi}_{n, k,m,l} x_{k, l}\right|^2 \right]\nonumber\\
 &\approx  \frac{2  \pi^2 \cos(\phi)^2 A^2 \Pav}{\sigma_w^2} \sum_{b=1}^B  \| \Um_{b}^\H  \av(\phi)\av^\H (\phi)\odot \Bm \fv \|^2 \sum_{n=0}^{N-1}\sum_{m=0}^{M-1}   
\sum_{k=0}^{L-1} \sum_{l=0}^{M-1} \left|  \bar{\Psi}_{n, k,m,l} \right|^2 \\
I_{\tau \tau} 
%&= \frac{8\pi^2 A_0^2 (\Delta f)^2}{N^2M^2\sigma_w^2} \sum_{b=1}^B  \| \Um_{b}^\H \av(\phi)\av^\H (\phi)\fv %\|^2\sum_{n=0}^{N-1}\sum_{m=0}^{M-1}   
%\EE \left[\left|\sum_{k=0}^{L-1} \sum_{l=0}^{M-1}  \onev_N^\T \alphav_{n, k}(\nu_p)  \cv_M^\T\betav_{k, m, l} (\tau_p, \nu_p) 
%x_{k, l}\right|^2\right] \nonumber\\
&\approx  \frac{4\pi^2 A^2 (\Delta f)^2 \Pav}{N^2M^2}  \| \Um_{b}^\H \av(\phi)\av^\H (\phi)\fv \|^2
\sum_{k=0}^{L-1} \sum_{l=0}^{M-1}  \left|\onev_N^\T \alphav_{n, k}(\nu)  \cv_M^\T\betav_{k, m, l} (\nu, \tau) 
\right|^2 \\
I_{\nu \nu} 
%&=-   \frac{8\pi^2 A_0^2}{N^2M^2\sigma_w^2} \sum_{b=1}^B  \| \Um_{b}^\H \av(\phi)\av^\H (\phi)\fv \|^2 \sum_{n=0}^{N-1}\sum_{m=0}^{M-1}    %\EE  \left[\left|\sum_{k=0}^{L-1} \sum_{l=0}^{M-1} d_{k, l} (\tau_0, \nu_0) x_{k, l}\right|^2\right]\nonumber\\
&\approx -\frac{4\pi^2 A^2 \Pav}{N^2M^2} \sum_{b=1}^B  \| \Um_{b}^\H \av(\phi)\av^\H (\phi)\fv \|^2 \sum_{n=0}^{N-1}\sum_{m=0}^{M-1}\sum_{k=0}^{L-1} \sum_{l=0}^{M-1} \left| d_{k, l} (\nu, \tau) \right|^2
\end{align}
Next, we derive the off-diagonal elements. First we remark
\begin{align}
I_{A \psi} &=\frac{2}{\sigma_w^2} \Re\left\{  j A  \sum_{b=1}^B \|\Um_{b}^\H \av(\phi)\av^\H (\phi)\fv \|^2 \sum_{n=0}^{N-1}\sum_{m=0}^{M-1}\EE\left[ \left|\sum_{k=0}^{L-1} \sum_{l=0}^{M-1}
\bar{\Psi}_{n, k,m,l}x_b[k, l]\right|^2 \right]\right\} \nonumber\\
%&= \frac{2}{\sigma_w^2} \Re\left\{  j A_0  \sum_{b=1}^B \|\Um_{b}^\H \av(\phi_0)\av^\H (\phi_0)\fv \|^2 \sum_{n=0}^{N-1}\sum_{m=0}^{M-1} \left|\sum_{k=0}^{L-1} \sum_{l=0}^{M-1}
%\bar{\Psi}_{n, k,m,l}x_{k, l}\right|^2 \right\} \nonumber\\
&=0 
 \end{align}
 By using  \eqref{eq:trace} and \eqref{eq:iidsymbol},  we have
 \begin{align}
  I_{ \phi A} 
  %& =  \frac{2}{\sigma_w^2} \Re\left\{\sum_{b=1}^B \sum_{n=0}^{N-1}\sum_{m=0}^{M-1} j \pi A_0 \cos(\phi_0)
%\| \Um_b^\H \av(\phi_p)\|^2 \fv^\H \av(\phi_p)  \av^\H(\phi_0) \odot \Bm \fv 
% \EE[ \left|\sum_{k=0}^{L-1} \sum_{l=0}^{M-1}\bar{\Psi}_{n, k,m,l}x_b[k, l] \right|^2] \right\}  \nonumber \\ 
 &\approx   - \frac{2\pi A \cos(\phi) \Pav}{\sigma_w^2} \sum_{b=1}^B \| \Um_b^\H \av(\phi)\|^2\Im\{ \fv^\H \av(\phi)  \av^\H(\phi) \odot \Bm \fv\}  \sum_{n=0}^{N-1}\sum_{m=0}^{M-1} \left|  \bar{\Psi}_{n, k,m,l} \right|^2\\
  I_{\phi  \tau}  &\approx     \frac{4\pi \Delta f \pi \cos(\phi) A^2 \Pav}{NM\sigma_w^2}  \sum_{b=1}^B  \| \Um_b^\H \av(\phi)\|^2 
\sum_{n=0}^{N-1}\sum_{m=0}^{M-1}    \sum_{k=0}^{L-1} \sum_{l=0}^{M-1}
\Re\{ \fv^\H \av(\phi)  \av^\H(\phi) \odot \nonumber \\
& \;\;\;\;\;\;\;\;\;\;\;\;\Bm \fv \bar{\Psi}^ * _{n, k,m,l}\onev_N^\T \alphav_{n, k}(\nu)  \cv_M^\T\betav_{k, m, l} (\nu_, \tau) \} \\
   I_{\phi  \nu} &\approx   \frac{4\pi^2 \cos(\phi) A^2\Pav}{NM\sigma_w^2}  \sum_{b=1}^B  \| \Um_b^\H \av(\phi)\|^2  \sum_{n=0}^{N-1}\sum_{m=0}^{M-1}
    \sum_{k=0}^{L-1} \sum_{l=0}^{M-1}\Re\{ \fv^\H \av(\phi)  \av^\H(\phi) \odot \Bm \fv 
 \bar{\Psi}^*_{n, k,m,l}d_{k, l} (\nu, \tau) \}\\
  I_{\phi  \psi} &\approx \frac{2  \pi \cos(\phi) A^2 \Pav}{\sigma_w^2}  \sum_{b=1}^B\| \Um_b^\H \av(\phi)\|^2 \Re\{\fv^\H \av(\phi)  \av^\H(\phi) \odot \Bm \fv  \}
  \sum_{n=0}^{N-1}\sum_{m=0}^{M-1}  \sum_{k=0}^{L-1} \sum_{l=0}^{M-1} \left|\bar{\Psi}_{n, k,m,l} \right|^2 
\end{align} 
Similarly, we have 
 \begin{align}
 I_{A \tau}
 %&= \frac{2}{\sigma_w^2} \Re\left\{ \sum_{b=1}^B\sum_{t=1}^{\Nrf} \sum_{n=0}^{N-1}\sum_{m=0}^{M-1} \left[\frac{\partial s_{b,p}[n ,m, t]}{\partial A_p}\right]^* %\left[\frac{\partial s_{b,p}[n, m,t]}{\partial \tau_p}\right]\right\} \nonumber\\
% &= \frac{2}{\sigma_w^2} \Re\left\{ \sum_{b=1}^B \sum_{n=0}^{N-1}\sum_{m=0}^{M-1} \frac{j2\pi \Delta f}{NM}A_p
%|\Um_{b}^\H \av(\phi_p)\av^\H (\phi_p)\fv|^2    
% \left(\sum_{k,l} \bar{\Psi}^p_{n, km,l} x_{k, l} \right)^* \left(\sum_{k',l'}x_{k', l'} 
% \sum_{n'} \alpha_{n, k',n'} \sum_{m'} m' \beta_{k', m, l',m'}\right)
%  \right\} \nonumber\\
  &=\frac{2}{\sigma_w^2} \Re\Bigg\{\EE \sum_{b=1}^B \sum_{n=0}^{N-1}\sum_{m=0}^{M-1} \frac{j2\pi \Delta f}{NM}A_p
\|\Um_{b}^\H \av(\phi)\av^\H (\phi)\fv\|^2    
 \left(\sum_{k,l} x_b[k, l]   \onev_N^\T \alphav_{n, k} \onev_M^\T \betav_{k, m, l} x_{k, l} \right)^* \nonumber\\
 & \;\;\;\;\;\; \left(\sum_{k',l'}x_b[k', l'] 
\onev_N^\T \alphav_{n, k'} \cv_M^\T \betav_{k', m, l'}\right)
  \Bigg\} \nonumber\\ 
  &\approx  \frac{2}{\sigma_w^2} \Re\left\{ \sum_{b=1}^B\sum_{n=0}^{N-1}\sum_{m=0}^{M-1} \frac{j2\pi \Delta f}{NM}A
\|\Um_{b}^\H \av(\phi)\av^\H (\phi)\fv\|^2    
\sum_{k,l}\EE[ |x_b[k, l]|^2]  |\onev_N^\T \alphav_{n, k}|^2 (\onev_M^\T \betav_{k, m, l} )^*  \cv_M^\T \betav_{k, m, l}%( \sum_{n'} \alpha_{n, k,n'} \sum_{m'} \beta_{k, m, l,m'})^*\sum_{n''} \alpha_{n, k,n''} \sum_{m''} m'' \beta_{k, m, l,m''}
  \right\} \nonumber  \\
  &=-
   \frac{4\pi \Delta f A \Pav}{\sigma_w^2 NM}\sum_{b=1}^B\sum_{n=0}^{N-1}\sum_{m=0}^{M-1}
\|\Um_{b}^\H \av(\phi)\av^\H (\phi)\fv\|^2    
\sum_{k,l} |\onev_N^\T \alphav_{n, k}(\nu) |^2  \Im\left\{ (\onev_M^\T \betav_{k, m, l} (\tau))^*  (\cv_M^\T \betav_{k, m, l})\right\} \\
  I_{A \nu} &= \frac{2}{\sigma_w^2} \Re\left\{ \EE \sum_{b=1}^B \sum_{n=0}^{N-1}\sum_{m=0}^{M-1} 
 \frac{j2\pi}{NM}  A \|\Um_{b}^\H \av(\phi)\av^\H (\phi)\fv\|^2    
\sum_{k,l} \bar{\Psi}^*_{n, k,m,l} x^*_b[k, l]  \sum_{k', l'} x_b[k', l'] d_{k', l'}
\right\} \nonumber\\
&\approx \frac{2 \Pav}{\sigma_w^2} \Re\left\{ \sum_{b=1}^B\sum_{n=0}^{N-1}\sum_{m=0}^{M-1} \frac{j2\pi}{NM}  
 A\|\Um_{b}^\H \av(\phi)\av^\H (\phi)\fv\|^2    
\sum_{k,l} \bar{\Psi}^*_{n, k,m,l} d_{k, l}
\right\}\nonumber \\
&=-\frac{4\pi \Pav A}{\sigma_w^2 NM}   \sum_{b=1}^B\sum_{n=0}^{N-1}\sum_{m=0}^{M-1}  \|\Um_{b}^\H \av(\phi)\av^\H (\phi)\fv\|^2    
\Im\left\{ \sum_{k,l} \bar{\Psi}^*_{n, k,m,l} d_{k, l} \right\}
\end{align}
\begin{align}
 I_{ \tau \nu} &\approx  \frac{8\pi^2 A^2 \Delta f \Pav}{(NM)^2\sigma_w^2}   \sum_{b=1}^B   \|\Um_{b}^\H \av(\phi)\av^\H (\phi)\fv \|^2
\sum_{n=0}^{N-1}\sum_{m=0}^{M-1}\sum_{k=0}^{L-1} \sum_{l=0}^{M-1} \Re\{ (\onev_N^\T \alphav_{n, k}(\nu)  \cv_M^\T\betav_{k, m, l} (\nu, \tau) )^* d_{k, l} (\nu, \tau)\} 
\\
 I_{\tau  \psi} &\approx  \frac{2}{\sigma_w^2}\frac{2\pi \Delta f   A^2 \Pav}{NM} \|\Um_{b}^\H \av(\phi)\av^\H (\phi)\fv \|^2
\sum_{n=0}^{N-1}\sum_{m=0}^{M-1}\sum_{k=0}^{L-1} \sum_{l=0}^{M-1} \Re\{( \onev_N^\T \alphav_{n, k}(\nu)  \cv_M^\T\betav_{k, m, l} (\nu, \tau) )^*\bar{\Psi}_{n, k,m,l}\}
  \\
  I_{\tau A} &\approx -  \frac{2}{\sigma_w^2}\frac{2\pi \Delta f  A \Pav}{NM} \sum_{b=1}^B    \|\Um_{b}^\H \av(\phi)\av^\H (\phi)\fv \|^2
\sum_{n=0}^{N-1}\sum_{m=0}^{M-1}\sum_{k=0}^{L-1} \sum_{l=0}^{M-1}\Im\{ (\onev_N^\T \alphav_{n, k}(\nu)  \cv_M^\T\betav_{k, m, l} (\nu, \tau))^* \bar{\Psi}_{n, k,m,l}\}\\
 I_{\nu \psi} &\approx \frac{2}{\sigma_w^2} \frac{2\pi A_0^2 \Pav}{NM} \sum_{b=1}^B  \| \Um_{b}^\H \av(\phi)\av^\H (\phi)\fv\|^2\sum_{n=0}^{N-1}\sum_{m=0}^{M-1}\sum_{k=0}^{L-1} \sum_{l=0}^{M-1}  \Re\{d^*_{k, l} (\nu, \tau)  \bar{\Psi}_{n, k,m,l}\}
\end{align}

\bibliography{IEEEabrv,book}

\end{document}